\documentclass[aps,prx,twocolumn,floatfix]{revtex4}
\usepackage{bm}
\usepackage{epsf}
\usepackage{amssymb}
\usepackage{amsmath}
\usepackage{graphicx}
\usepackage{rotating}
\usepackage{epsfig}
\usepackage{psfrag}
\usepackage{amsmath}
\usepackage{hyperref}
\usepackage{subfigure}
\usepackage{braket}
\usepackage{tikz}
\usepackage{stmaryrd}
\usepackage{wasysym}
\usepackage{amsfonts}
\usepackage{soul}

\newcommand{\enni}{\noindent}

\begin{document}

\title{Kitaev spin-orbital bilayers and their moir\'e superlattices}

\author{Emilian Nica$^1$, Muhammad Akram$^{1,2}$, Aayush Vijayvargia$^1$, Roderich Moessner$^3$, Onur Erten$^1$}
\affiliation{$^1$Department of Physics, Arizona State University, Tempe, AZ 85287, USA\\
$^2$Department of Physics, Balochistan University of Information Technology, Engineering and Management Sciences (BUITEMS), Quetta 87300, Pakistan\\ $^3$Max-Planck-Institut f\"ur Physik komplexer Systeme, N\"othnitzer Strasse 38, 01187 Dresden, Germany}

\begin{abstract}
We determine the phase diagram of a bilayer, Yao-Lee spin-orbital model with inter-layer interactions ($J$), for several stackings and moir\'e superlattices. For AA stacking, a gapped $\mathbb{Z}_{2}$ quantum spin liquid phase emerges at a finite $J_{c}$. We show that this phase survives in the well-controlled large-$J$ limit, where an isotropic  honeycomb toric code emerges. For moir\'e superlattices, a finite-$\mathbf{q}$ inter-layer hybridization is stabilized. This connects inequivalent Dirac points, effectively `untwisting' the system. Our study thus provides insight into the spin-liquid phases of bilayer spin-orbital Kitaev materials.
\end{abstract}
\maketitle

Quantum spin liquids (QSLs) are disordered phases of magnetic systems with emergent exotic properties arising from their underlying topological character~\cite{ Balents_Nature2010, Zhou_RMP2017, Wen_RMP2017, Knolle_AnnRevCondMatPhys2019, Broholm_Science2020}. The Kitaev model on the honeycomb lattice~\cite{Kitaev_AnnPhys2006,Hermanns_2018} is of particular significance as the first member of a family of exactly-solvable models. Recent years witnessed experimental progress in identifying candidate materials which include a number of iridates~\cite{HwanChun_NatPhys2015} and $\alpha$-RuCl$_3$~\cite{Takagi_NatRevPhys2019}. Kitaev interactions can also be strong in other van der Waals (vdW) materials such as CrI$_3$~\cite{Lee_PRL2020, Blei_APR2021}. Moreover, vdW materials can be arranged in stacking patterns and twisted to form moir\'e superlattices, potentially leading to new phases. Indeed, recent theoretical studies~\cite{Tong_ACS2018, Hejazi_PNAS2020, Hejazi_PRB2021, Akram_PRB2021, Tong_PRR2021, Akram_NanoLett2021} predict several magnetic phases in twisted vdW magnets, partially realized experimentally~\cite{Xu_NatNano2021, Song_Science2021}.


We study the zero-temperature phase diagram of bilayer versions of Kitaev spin-orbital models, initially proposed by Yao and Lee~\cite{Yao_PRL2011}, with additional inter-layer Heisenberg interactions. Spin-orbital models are generalizations of the original Kitaev model with extra local orbital degrees of freedom (DOF) and Kugel-Khomskii interactions for 
spin and orbital sectors~\cite{Kugel_SovPhys1982}, \cite{Yao_PRL2009, Wang_PRB2009, Wu_PRB2009, Yao_PRL2011, Carvalho_PRB2018, Seifert_PRL2020, Chulliparambil_PRB2020, Natori_PRL2020, Chulliparambil_PRB2021, Tsvelik_arxiv2021}. Much like Kitaev's original proposal, spin and orbital DOF can each be represented in terms of three-flavored sets of Majorana fermions. The Yao-Lee model exhibits an emergent $\mathbb{Z}_{2}$ gauge symmetry with gapped flux excitations (visons) defined exclusively in terms of the orbital DOF~\cite{Yao_PRL2011}. The inter-layer spin-exchange interactions commute with the intra-layer flux operators, in contrast to the Kitaev model and subsequent bilayer realizations~\cite{Seifert_PRb2018, Tomishige_PRB2018, Tomishige_PRB2019, May-Mann_PRB2020}.  
We take advantage of this unique feature by considering only the lowest-energy, zero-flux sector. Furthermore, we treat the spin-exchange interactions in the Hartree approximation. This introduces an effective inter-layer hybridization for the itinerant Majorana fermions associated with the spin DOF. A non-zero expectation value indicates the formation of inter-layer spin-singlets, as shown in the Supplemental Material (SM). The conservation of the fluxes in the Yao-Lee bilayer, which are defined exclusively in terms the orbital DOF, stands in clear contrast to bilayers based on Kitaev's original model. As shown below, this leads to distinct phase diagrams and to an enhanced stability of topological QSL phases in Yao-Lee bilayers.

\begin{figure}[t!]
\includegraphics[width=\columnwidth]{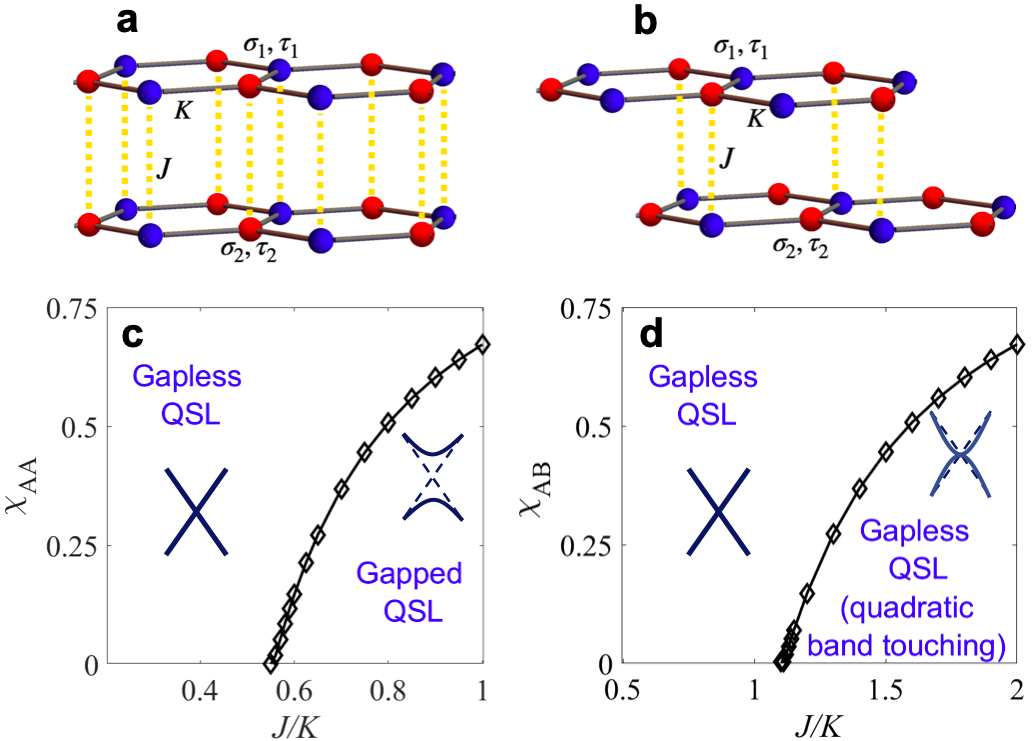}
\caption{Illustration of Yao-Lee bilayer model for (a) AA and (b) AB stacking patterns. $K$ and $J$ are the intra-layer Kitaev and inter-layer Heisenberg exchange terms, respectively. (c) Effective inter-layer hybridization for AA stacking. Finite $\braket{\chi_{\rm{AA}}}$ indicates the formation of inter-layer singlets and leads to gapped itinerant Majorana fermions. (d) Same for AB stacking, leading to quadratic band touching.}
\label{Fig:1}
\end{figure}

We focus on AA stacking and moir\'e superlattices, which exhibit fully-gapped spectra, but also briefly cover the gapless, AB stacking case. For AA stacking, the effective hybridization becomes non-zero at a finite value of the inter-layer exchange coupling, and opens a gap in the itinerant Majorana fermion spectrum. This signals a topological phase transition to a gapped $\mathbb{Z}_{2}$ QSL. We support our Hartree approximation with two additional considerations. First, we show that the bilayer model is equivalent to an attractive Hubbard model with three flavors of complex fermions, for our choice of gauge. Previous quantum Monte Carlo (QMC) studies have shown that the Hubbard model exhibits a single transition to a 
charge density wave (CDW) phase~\cite{Xu_arXiv2019}, which is equivalent to the bilayer with a non-zero inter-layer hybridization. Secondly, we show that in the limit of large inter-layer 
interactions, the bilayer model maps onto Kitaev's toric code~\cite{Kitaev_AnnPhys2003}, which is gapped and exhibits topological order. This naturally 
suggests that the gapped phase predicted by the Hartree approximation is adiabatically connected to the toric code. However, first-order transitions, possibly involving changes in the flux configurations, cannot be completely excluded. For AB stacking, the formation of inter-layer spin singlets leaves the itinerant Majorana fermions gapless with quadratic band touching, in analogy with bilayer graphene~\cite{Rozhkov_PhysRep2016}. For moir\'e superlattices, we consider both uniform ($\mathbf{q}=0$) and modulated inter-layer effective hybridizations ($\mathbf{q} \neq 0$). In contrast to the $\mathbf{q}=0$ case, the finite-$\mathbf{q}$ hybridization connects inequivalent Dirac points, effectively `untwisting' the system, and opening a gap. This leads to the emergence of a gapped $\mathbb{Z}_{2}$ QSL, as for AA stacking. 

Kitaev spin-orbital models can be realized in spin-orbit coupled $4d$ and $5d$ Mott insulators, as predicted by several recent studies~\cite{Natori_PRB2019, Seifert_PRL2020, Xu_PRL2020, Stavropoulos_PRL2019}. For instance, an enhanced SU(4) symmetry~\cite{Yamada_PRL2018} has been advanced for $\alpha$-ZrCl$_3$. 

{\sl Model.} 
Our models include intra-layer Yao-Lee~\cite{Yao_PRL2011} interactions 
on a honeycomb lattice
($H_{\nu}$),
and inter-layer, antiferromagnetic Heisenberg interactions ($H_{\rm{I}}$):

\noindent \begin{align}
H= H_{\nu} + H_{\rm{I}}
\end{align}

\noindent \begin{align}
    H_{\nu}= \sum_{\alpha {\rm -links}, \langle ij \rangle} K^{(\alpha)} \left(\tau_{\nu, i}^{(\alpha)} \tau_{\nu,j}^{(\alpha)}\right) 
\left( \pmb{\sigma}_{\nu,i} \cdot \pmb{\sigma}_{\nu,j} \right)
\label{Eq:YL}
\end{align}

\noindent \begin{align}
H_{\text{I}}=\sum_{ij} J_{ij} \pmb{\sigma}_{1i}\cdot \pmb{\sigma}_{2j}.
\end{align}

\noindent We first focus on $H_{\nu}$, where $K^{(\alpha)}$ is the nearest neighbor coupling (NN) constant for type-$\alpha$ links ($\alpha \in \{x,y,z\}$) (Fig~\ref{Fig:1}~(a)-(b)). The lattice sites are labeled by $i$ and $j$, while $\nu \in \{1,2\}$ denotes the two layers.  An exact solution is obtained by introducing Majorana fermion representations for the spin and orbital DOF in each layer:  $\sigma_{\nu,j}^{(\alpha)}=-i\epsilon^{\alpha \beta \gamma}c_{\nu,j}^{(\beta)}c_{\nu,j}^{(\gamma)}/2$ and $\tau_{\nu,j}^{(\alpha)}=-i\epsilon^{\alpha \beta \gamma}d_{\nu,j}^{(\beta)}d_{\nu,j}^{(\gamma)}/2$~\cite{Yao_PRL2011}. Note that we use a normalization convention for the Majorana fermions where $\{c^{(\alpha)}_{\mu, i}, c^{(\beta)}_{\nu, j} \} = 2 \delta_{\alpha \beta} \delta_{\mu\nu} \delta_{ij}$, and similarly for the $b$'s. These representations are redundant and the physical states in each layer must be restricted to the eigenstates of $D_{\nu,i}=-i c_{\nu,i}^{(x)} c_{\nu,i}^{(y)} c_{\nu,i}^{(z)} d_{\nu,i}^{(x)} d_{\nu,i}^{(y)} d_{\nu,i}^{(z)}$ operators with eigenvalues $1$. As in Kitaev's model, these constraints can be imposed via projection operators $P_{\nu}=\prod_i (1+D_{\nu,i})/2$. The intra-layer Hamiltonians in the Majorana representation can be expressed as $H_{\nu}=  P_{\nu} \mathcal{H}_{\nu} P_{\nu}$, where

\noindent \begin{align}
    \mathcal{H}_{\nu}=\sum_{\langle ij \rangle} K^{(\alpha)} u^{\alpha}_{\nu, ij}[ic_{\nu,i}^{(x)} c_{\nu,j}^{(x)}+i c_{\nu,i}^{(y)} c_{\nu,j}^{(y)} + ic_{\nu,i}^{(z)} c_{\nu,j}^{(z)}].
\end{align}

\noindent The bond operators $u^{(\alpha)}_{\nu, ij}=-id_{\nu,i}^{(\alpha)}d_{\nu,j}^{(\alpha)}$, where $i,j$ are on the A and B sublattices, respectively, commute with $\mathcal{H}_{\nu}$, and are therefore conserved with eigenvalues $\pm 1$. Both $\mathcal{H}_{\nu}$ are invariant under separate $\mathbb{Z}_{2}$ gauge transformations $c_{\nu, i}^{(\alpha)} \rightarrow -c_{\nu,i}^{(\alpha)};$ $u^{(\alpha)}_{ij} \rightarrow -u^{(\alpha)}_{ij}$ with flux operators which are defined by the product of the $u^{(\alpha)}_{ij}$ around hexagonal plaquettes. 

Lieb's theorem~\cite{Lieb_PRL1994} predicts that the ground state lies in the zero-flux sector, with a finite vison gap. We can obtain the itinerant Majorana spectrum by choosing a gauge where $u_{ij}=1 \ \forall \ \langle ij \rangle$ in both layers. Unless otherwise stated, we use this choice throughout. 
The three flavors of itinerant Majorana fermions have identical spectra which are gapless for $K_x+K_y>K_z$. We consider the symmetric, gapless case with $K_x=K_y=K_z=K$.

We now consider the inter-layer interactions in $H_{\rm{I}}$. Unlike in the Kitaev model, the visons in the Yao-Lee model are defined exclusively in terms of the orbital DOF, while the itinerant Majorana excitations stem from the spin DOF alone. Consequently, additional terms involving the spin DOF only, including a bilayer coupling, commute with the flux operators. The resulting spectrum can \emph{a priori} preserve the gapped flux excitations, in contrast to the original Kitaev model~\cite{Seifert_PRb2018, Tomishige_PRB2019, May-Mann_PRB2020}. Consequently, we consider Yao-Lee bilayers coupled via inter-layer antiferromagnetic Heisenberg interactions in $H_{\rm{I}}$.
Note that we allow for general inter-layer $J_{ij}$ coupling beyond NN. The bilayer Hamiltonian in the Majorana representation is 
$\mathcal{H}= \sum_{\nu} \mathcal{H}_{\nu} +\mathcal{H}_{\text{I}}$ where

\noindent \begin{align}
    \mathcal{H}_{\text{I}}= \sum_{i,j;\alpha \neq \beta} \frac{J_{ij}}{2}\left( c_{1i}^{(\alpha)}c_{2j}^{(\alpha)}c_{1i}^{(\beta)}c_{2j}^{(\beta)} \right).
    \label{eq:fullH}
\end{align}

{\sl Self-consistent solutions for AA and AB stacking patterns.} The inter-layer interactions are bi-quadratic in the itinerant Majorana operators, thus precluding a closed-form solution. Instead, we treat the inter-layer interactions within a Hartree approximation. This approach is supported by additional considerations, as discussed below. 

For the purpose of illustration, we restrict the inter-layer coupling to NN pairs. We do not expect that weaker couplings beyond NNs will alter our conclusions. The on-site mean-field (MF) parameters 
$\langle \chi^{(\alpha)}_i \rangle = \langle ic_{1i}^{(\alpha)} c_{2i}^{(\alpha)}\rangle$ preserve the SO(3) spin symmetry, and, 
we drop the corresponding flavor indices for most of the following discussion. 

Before proceeding with a detailed presentation of the results, we first clarify the nature of the MF parameters. In the absence of intra-layer Yao-Lee interactions ($K^{(\alpha)}=0$), the decoupled, inter-layer, spin-singlet states for overlapping sites can equally be described by two eigenstates of $ \chi^{(\alpha)}_{i}$, with eigenvalues $\pm 1$ for each $\alpha$, as shown Sec. I of the SM. The Ising-like nature of these states stems from a redundancy in the representation of the decoupled singlets in terms of the $c$ Majorana fermions. Once the intra-layer interactions are turned on, and a set of bond variables ($u^{(\alpha)}_{ij}$) is chosen, we obtain a unique MF solution with $\braket{ \chi^{(\alpha)}_{i}} \neq 0$, which is identical for the three flavors. These finite MF parameters likewise indicate the formation of inter-layer spin-singlets in the physical ground-state (GS). However, the Ising-like nature of these parameters is not immediately physical, since the non-trivial phases that we find are not described in terms of a local order parameter. We further elucidate these aspects in the following. 

As previously mentioned, we carry out the Hartree approximation in a gauge where all $u_{ij}=1$ in both layers, and obtain the GS 

\noindent \begin{align}
\ket{\Psi} = \ket{\forall~u^{(\alpha)}_{\nu, ij}=1} \otimes \ket{\braket{\chi^{(\alpha)}_{i} \neq 0}}.
\end{align}

\noindent Importantly, $\braket{\chi^{(\alpha)}_{i}}_{\Psi}$ is not a well-defined, Landau-Ginzburg order parameter for the bilayer. Indeed, any gauge transformation, implemented for instance by $D_{\nu, i} \ket{\Psi}$, changes the sign of the associated $\braket{\chi^{(\alpha)}_{i}}$ together with those of the three bonds extending from $i$ in layer $\nu$. Furthermore, the physical GS is obtained by applying the projector $P$ to $\ket{\Psi}$ as

\noindent \begin{align}
\ket{\Psi}_{\rm{Phys}} = & P \ket{\Psi}.
\end{align}

\noindent $\ket{\Psi}_{\rm{Phys}}$  amounts to a linear superposition of all gauge-symmetrized states which preserve a net zero flux, as shown in Sec. II A of the SM. States with finite $\pm |\braket{\chi^{(\alpha)}_{i}}|$ occur with equal weight, implying that $\braket{\chi^{(\alpha)}_{i}}_{\rm{Phys}}=0$. 

In order to characterize transitions in the physical GS, we instead consider a gauge-invariant correlator

\noindent \begin{align}
\braket{C^{(\alpha)}_{ij}}_{\rm{Phys}} = & 
\Bigg \langle 
\left( \prod^{'}_{\alpha-\text{links},\braket{i'j'}} u^{(\alpha)}_{1, i'j'} u^{(\alpha)}_{2, i'j'} \right) \chi^{(\alpha)}_{i} \chi^{(\alpha)}_{j}
\Bigg \rangle_{\rm{Phys}},
\end{align}   

\noindent where the strings of bonds connect operators at the end sites $i,j$. In Sec. II B of the SM, we show that the expectation value of $C^{(\alpha)}_{ij}$ in the physical GS matches that of a two-point correlator for $\chi^{(\alpha)}$ in $\ket{\Psi}$. 

\noindent \begin{align}
\braket{C^{(\alpha)}_{ij}}_{\rm{Phys}} = \braket{\chi^{(\alpha)}_{i}\chi^{(\alpha)}_{j}}_{\Psi}.
\label{Eq:GICr}
\end{align}

\noindent From this expression, long-range order in $\ket{\Psi}$ is equivalent to $\braket{\chi^{(\alpha)}_{i}}_{\Psi} \neq 0$. It follows that non-vanishing MF parameters imply a finite $\braket{C^{(\alpha)}_{ij}}_{\rm{Phys}}$, in the limit of infinite separation.
In Sec. II C of the SM, we express $\braket{C^{(\alpha)}_{ij}}_{\rm{Phys}}$ in terms of the spin  and orbital operators of the bilayer, and show that it signals  a topological phase transition to a gapped, $\mathbb{Z}_{2}$ QSL for the AA-stacked case, which involves the formation of inter-layer spin-singlets. We note that all subsequent conclusions regarding the MF parameters, obtained in the Hartree approximation and in a fixed gauge, are to be understood in the present context.

We now discuss our results in the Hartree approximation. For the AA stacking pattern, the A and B sublattice sites overlap (Fig.~\ref{Fig:1}~(a)). The inter-layer interactions involve two pairs of sites per unit cell: $\mathcal{H}_{\text{I}}=-2J[ \sum_{i \in A,\alpha} \langle \chi_{\rm{AA}} \rangle (ic_{1i}^{\alpha}c_{2i}^{\alpha})+\sum_{j \in B,\alpha} \langle \chi_{\rm{BB}}\rangle(ic_{1j}^{\alpha}c_{2j}^{\alpha})]$. Solutions which are both uniform and symmetric in the sublattice index ($\braket{\chi_{\rm{AA}}}=\braket{\chi_{\rm{BB}}}$) amount to gapless itinerant Majorana fermions, with shifted Dirac cones. In contrast, when the hybridization has an alternating sign on the two sublattices ($\langle \chi_{\rm{AA}} \rangle = -\langle \chi_{\rm{BB}} \rangle$), the spectrum is gapped, leading to a lower ground-state energy. Our self-consistent solutions are shown in Fig.~\ref{Fig:1}~(c) as functions of $J/K$. We find that the critical value for this transition is $J_c/K=0.55$. 

To establish the stability of our solutions beyond the Hartree approximation, we map $\mathcal{H}$ to an equivalent form by using complex fermions $f_{i}^{\alpha}=(c_{1i}^{\alpha}+ic_{2i}^{\alpha})/2$: 

\noindent \begin{align}
     \mathcal{H} =2K \sum_{\langle ij \rangle, \alpha}(if_{A,i}^{\alpha \dagger}f_{B,j}^{\alpha} +{\rm H.c.}) -2J\sum_{i}\left(n_i-\frac{3}{2}\right)^2,
     \label{eq:Hcf}
\end{align}

\noindent where $n_i=\sum_{\alpha}f_i^{\alpha\dag}f_i^{\alpha}$. For $J>0$, eq. \ref{eq:Hcf} describes an attractive Hubbard model with three flavors of complex fermions. This model exhibits a single, broken-symmetry CDW phase with finite $\langle n_{\rm{A}} \rangle = - \langle n_{\rm{B}} \rangle$, as determined by QMC~\cite{Xu_arXiv2019}. This Ising order parameter acts as a mass term for the complex fermions, and gaps their spectrum. It is equivalent to a solution in which  $\braket{\chi}$ alternates between sublattices in the Yao-Lee bilayer. Importantly, the Hubbard model and CDW order parameter were obtained by fixing the gauge. While the CDW breaks inversion symmetry in the Hubbard model, the same cannot be said 
of the physical GS of the bilayer model. As previously mentioned, the order parameters obtained in a fixed gauge are physically meaningful only in relation to the gauge-invariant correlator $\braket{C^{(\alpha)}_{ij}}_{\rm{Phys}}$ defined in Eq.~\ref{Eq:GICr}. 

The GS obtained in the Hartree approximation for AA stacking has a fourfold topological degeneracy, as shown in Sec. IX of the SM. This result is corroborated by the perturbative analysis in the large-$J$ limit discussed in the following.  

For AB stacking, the A sublattice sites of layer 1 lie directly on top of the B sublattice sites of layer 2, with a single bond per unit cell, (Fig.~\ref{Fig:1}~(b)). Therefore, for finite $\braket{\chi}$ beyond $J_c/K \simeq 1.1$, the itinerant Majorana spectrum is similar to that of AB-stacked bilayer graphene with quadratic band touching~\cite{Rozhkov_PhysRep2016}. The self-consistent solutions for $\langle \chi_{\rm{AB}}\rangle$ are shown in Fig.~\ref{Fig:1}~(d). A mapping to an equivalent model as in the AA case is not apparent here. 

The choice of uniform $u_{ij}=1$ for both layers implies that the system persists in a zero-flux sector. This is supported  by additional MF calculations with several distinct non-zero flux patterns (Sec. III of the  SM), which indicate that the zero-flux states are lower in energy. Furthermore, the effective Hamiltonian in the large-$J$ limit (see below) similarly prefers this configuration.

We comment on the stability of the phases obtained in the Hartree approximation in the presence of additional inter-layer, NN, spin-exchange  interactions, which we realistically expect to be subleading. For the AA-stacked bilayer, the gapped phase obtained for $J>J_{c}$ is stable with respect to additional, infinitesimal, NN interactions. For the AB-stacked bilayer with $J>J_{c}$, our Hartree approximation predicts quadratic band touching, which implies a finite density of states for the itinerant Majorana fermions at zero energy. Additional inter-layer, NN interactions are therefore likely relevant in a renormalization-group sense. 
Establishing the nature of the low-energy phases in these cases requires further analysis, at Hartree level and beyond, and we reserve such questions for future study.

The GSs obtained in the fixed gauge survive projection onto the physical sector, as shown in Sec. II A of the SM.

\begin{figure}[t]
\includegraphics[width=\columnwidth]
{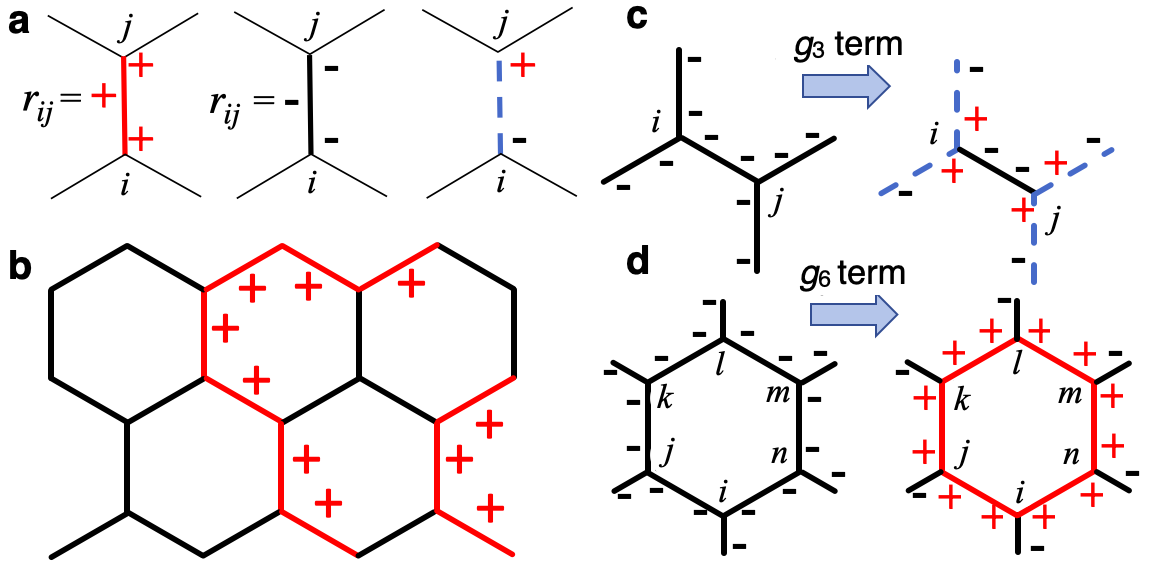}
\caption{(a) 
The bond operator $r_{ij}= \text{sgn}(p^{(\alpha)}_{i})=\text{sgn}(p^{(\alpha)}_{j})$,
which is defined in the ground-state manifold of the $g_{2}$ terms in Eq.~\ref{Eq:Hmlt_orbtl}. Red (black) solid lines correspond to $+~(-)$ bonds. Conversely, bond configurations  which include NNs with $\text{sgn}(p^{(\alpha)}_{i}) \neq\text{sgn}(p^{(\alpha)}_{j})$, shown here with blue dashed lines, are not labeled by $r_{ij}$ bonds. These configurations correspond to excited states. (b) A ground-state manifold configuration which minimizes the $g_2$ terms and which also obeys the local product constraint, equivalent to an Ising Gauss's law. (c) $g_3$ terms on sites $i,j$ flip four adjacent bonds. (d) $g_6$ terms flip the plaquette bond configurations. 
}
\label{Fig2}
\end{figure}

{\sl Limit of large inter-layer interactions with AA stacking pattern.} We consider the AA stacking pattern in the limit of large $J/K$. To zeroth order in the intra-layer ($K$) terms, the GS manifold consists of a collection of independent inter-layer spin singlets with degenerate orbital states. We derive an effective Hamiltonian on the GS manifold, perturbatively up to $6^{\rm th}$ order in $K/J$
\noindent \begin{eqnarray}
H_{\rm eff}&=& \sum_{\substack{\alpha {\rm -links} \\ \nonumber
\langle ij \rangle}} \left( g_2 \tau_{1i}^{(\alpha)} \tau_{2i}^{(\alpha)} \tau_{1j}^{(\alpha)} \tau_{2j}^{(\alpha)} + g_3 \sum_{\nu=1,2} \tau_{\nu i}^{(\alpha)} \tau_{\nu j}^{(\alpha)}\right) \\
&&+g_6\sum_{\hexagon ,\nu} W_p^\nu 
\label{Eq:Hmlt_orbtl}
\end{eqnarray}
where $W_p^{1(2)}$ is the flux operator defined on the honeycomb plaquettes on layer 1(2) as $W_p^{\nu} = \tau_{\nu i}^{(z)} \tau_{\nu j}^{(y)} \tau_{\nu k}^{(x)} \tau_{\nu l}^{(z)} \tau_{\nu m}^{(y)} \tau_{\nu n}^{(y)}$ (Fig.~\ref{Fig2}~(d)). Please consult Sections IV and V of the SM for additional details. The coupling constants are $g_2 =-K^2/4J$, $g_3 = -K^3/J^2$, and $g_6 = -K^6/(8J)^5$. The $g_2$ term describes Kitaev interactions around \emph{inter-layer plaquettes} while the $g_3$ term is a standard Kitaev interaction in each layer.
  
Note that $g_{6}$ terms promote uniform $W_{p}^{1(2)}=1$ corresponding to a zero-flux low-energy manifold. This configuration is preserved by the remaining terms which commute with the $W_{p}^{1(2)}=1$.

We first focus on the the leading $g_{2}$ terms, and define new operators $p_i^{(\alpha)} = \tau_{i1}^{(\alpha)}\tau_{2i}^{(\alpha)}$, which unlike the $\tau$'s, all commute with each other. Furthermore their product amounts to $p_i^{(x)}p_i^{(y)}p_i^{(z)} = -1$. Therefore, we use local basis states which are eigenstates of all $p$ operators and which also satisfy the product rule: $\{|-,-, -\rangle, |-,+, +\rangle, |+,-, +\rangle, |+,+, -\rangle \}$ where $\pm$ denotes the eigenvalue of $p^{(\alpha)}$, ($\alpha = x, y,z$). The $g_{2}<0$ terms favor equal-$p^{(\alpha)}$ states on NN sites. Therefore, in the GS manifold of the $g_2$ terms, it is possible to define bond variables $r_{ij} = \pm 1$ for pairs of $(+,+)$ and $(-,-)$ eigenvalues of $p^{(\alpha)}_{i/j}$, respectively. For configurations that do not minimize the $g_2$ terms, $r_{ij}$ is not defined (Fig. \ref{Fig2}~(a)). In addition to minimizing the $g_{2}$ terms, the GS manifold must also satisfy the local constraint due to $p_i^{(x)}p_i^{(y)}p_i^{(z)} = -1$. Taken together, these conditions are equivalent to bond configurations which obey an Ising Gauss's law $G_i^P = \prod_{\Ydown} r_{ij} = -1$ (Fig.~\ref{Fig2}~(b)). We stress that the $r_{ij}$ bond variables and Gauss's law are only defined in the GS manifold of the $g_{2}$ terms.

Next, we examine the effect of $g_3$ and $g_6$ terms acting on the GS manifold obtained from the combined effects of the $g_{2}$ terms and local product constraints. Each $\tau^{(\alpha)}_{1,2}$ acting on $|p_x, p_y, p_z\rangle$ preserves the corresponding $\alpha$ eigenvalues but flips the remaining two (see SM). Therefore, the $g_6$ terms acting on a plaquette flips all of the $r_{ij}$ bond variables therein 
(Fig~\ref{Fig2}~(d)), leading to an effective term
\begin{eqnarray}
-\kappa \sum_{\hexagon} \big( | \hexagon \rangle \langle \bar{\hexagon}| + {\rm H.c.} \big)
\label{eq:kappa}
\end{eqnarray}
In contrast, the single $g_3$ term on sites $\langle ij \rangle$, connects a ground-state configuration to excited states (Fig.~\ref{Fig2}~(c)). Consecutive application of $g_3$ terms around a plaquette leads to plaquette flips,  but these processes are subdominant with respect to those due to the $g_6$ term.

As shown in Sec. VI of the SM, the resonance term in eq. \ref{eq:kappa}, along with Gauss' law, describe Kitaev's toric code~\cite{Kitaev_AnnPhys2003} on a honeycomb lattice. We thus conclude that the bilayer model in the limit of large inter-layer spin exchange interactions is in a gapped abelian $\mathbb{Z}_{2}$ topological QSL phase. 

\begin{figure}[t]
\includegraphics[width=8cm]{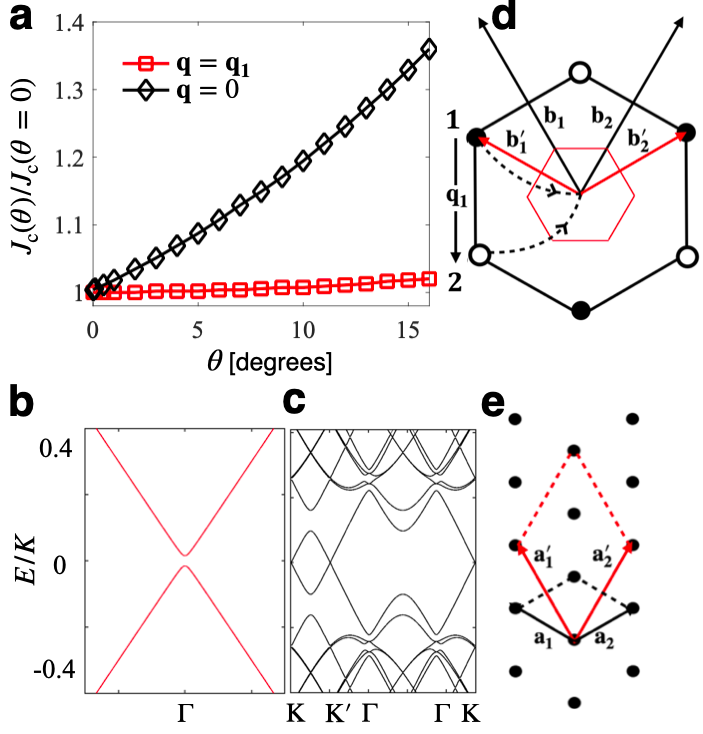}
\caption{Moir\'e superlattices: (a) critical inter-layer exchange ($J_c$) as a function of twisting angle $\theta$ for $\mathbf{q}=0$ and $\mathbf{q}=\mathbf{q}_1$. The spectrum for (b) $\mathbf{q}=\mathbf{q}_1$ and (c) $\mathbf{q}=0$. (d) The moir\'e lattice vectors (black) for $J<J_{c}$ and $J>J_{c}$ (red) corresponding to the finite-$\mathbf{q}$ solution. (e) Moir\'e reciprocal unit cells as in (d). $1$ and $2$ denote the Dirac points of the two layers which are separated by $\mathbf{q}_1$ in the absence of a finite inter-layer hybridization. When the latter acquires a finite value for $J>J_{c}$, the two Dirac points are shifted to the $\Gamma$ point of the folded BZ, and are subsequently gapped.
}
\label{Fig_twist}
\end{figure}

{\sl Self-consistent solutions for moir\'e superlattices}. We generalize the Hartree approximation to include the effects of small-angle twists. We follow Ref.~\citenum{Bistritzer_PNAS2011} to derive a low-energy theory defined on the moir\'e extended BZ, as shown in Sec. VII of the SM. 

To allow for non-vanishing inter-layer interactions under arbitrary, small twist angles, we extend the former beyond overlapping NN pairs and allow for an implicit decay with increasing pair separation. In general, this entails a decay of the Fourier components $J(\mathbf{k})$ with $|\mathbf{k}|$ 
, and involves interactions which are delocalized in the extended BZ.
In the low-energy limit, the interactions are naturally limited to the vicinity of a discrete set of equivalent crystal momenta throughout the extended BZ. In practice, we keep only $J(\mathbf{k})$ with $|\mathbf{k}|\lessapprox |2 \mathbf{K}_{00}|$, or twice the distance from the origin to the nearest Dirac point (Eq. S59 in the SM). We also assume that the retained Fourier components are all comparable in magnitude. The restrictions on the values of $J(\mathbf{k})$ allow us to explicitly consider the Yao-Lee bilayer analogs of flat bands in twisted bilayer graphene~\cite{Bistritzer_PNAS2011}. However, our conclusions are independent of this approximation, as discussed in the following. We also limit the hybridization to states in the vicinity of Dirac points in neighboring moir\'e reciprocal unit cells. This truncation is justified  in the low-energy limit, where small-momentum scattering processes are dominant.

The intra-layer terms amount to the usual Dirac fermions for the two layers, which are shifted with respect to each other due to twisting. The inter-layer interactions together with the approximations discussed previously can be written as

\noindent \begin{align}
\mathcal{H}_{\text{I}} = & - \frac{4J}{N} \sum_{nm} \left[ \braket{\chi^{\dag}_{00}(\mathbf{q})}
 \chi_{nm}(\mathbf{q}) 
+  \braket{\chi_{00}(\mathbf{q})}
 \chi_{nm}(-\mathbf{q})
 \right]
 \notag \\
 & + \text{H.c.}
 \label{Eq:HIMF}
\end{align}

\noindent where 

\noindent \begin{align}
\chi_{nm}(\mathbf{q})
= & i \sum_{\mathbf{k}}
\bigg[ 
c^{\dag, (\mu)}_{1, \alpha; nm}(\mathbf{k}) 
c^{(\mu)}_{2, \beta;nm}(\mathbf{k} -\mathbf{q})
\notag \\
& + e^{-i \pmb{\mathcal{G}}_{2} \cdot (\pmb{\tau}_{\alpha} - \pmb{\tau}_{\beta} )} c^{\dag, (\mu)}_{1, \alpha;n m}(\mathbf{k}) 
c^{(\mu)}_{2, \beta; n+1m}(\mathbf{k}-\mathbf{q}) 
\notag \\
& +  e^{-i \pmb{\mathcal{G}}_{3} \cdot (\pmb{\tau}_{\alpha} - \pmb{\tau}_{\beta} )} c^{\dag, (\mu)}_{1, \alpha;n m}(\mathbf{k}) 
c^{(\mu)}_{2, \beta; nm+1}(\mathbf{k}-\mathbf{q}) 
\bigg]
\label{Eq:TwHy}
\end{align}

\noindent , while 

\noindent \begin{align}
c^{\dag, (\mu)}_{\alpha,1/2}(\mathbf{K}_{00} + \mathbf{k} - n \mathbf{b_{2}} - m\mathbf{b}_{3}) = c^{\dag, (\mu)}_{\alpha,1/2; nm}(\mathbf{k})
\end{align}

\noindent are states with an effective Dirac dispersion which is shifted by the moir\'e reciprocal vectors $\mathbf{b}_{2,3}$ with respect to the Dirac point centered on the moir\'e first BZ at $n=m=0$. $\mathbf{K}_{00}$ is the position of the Dirac point of layer 1 in the first BZ while $\pmb{\mathcal{G}}_{2,3}$ are the reciprocal unit vectors of layer 1. The sums over momenta $\mathbf{k}$ cover the extended moir\'e BZ, with an implicit cutoff. The vectors $\pmb{\tau}_{\alpha, \beta}$ denote the shift of the A, B sublattices in layers 1 and 2, respectively. $\mathbf{q}$ is a vector contained within a single moir\'e reciprocal unit cell. As already mentioned, our approximations, and the cutoff for $J(\mathbf{k})$ in particular, ensure that the form of the effective hybridization in Eqs.~\ref{Eq:HIMF} and~\ref{Eq:TwHy} bears a close resemblance to that of twisted bilayer graphene~\cite{Bistritzer_PNAS2011}. For more details on the MF procedure, please see Sec. VII of the SM.

We consider two cases, one for $\mathbf{q}=0$ corresponding to a uniform inter-layer hybridization, and another for finite $\mathbf{q}=\mathbf{q}_{1}$ where $\mathbf{q}_{1}=-8\pi/3\sin(\theta/2)\hat{y}$~\cite{Bistritzer_PNAS2011} which denotes the shift between the Dirac points in layers 1 and 2 in the first BZ due to twisting (see Fig.~\ref{Fig_twist}~(e)). In both cases, $\langle \chi_{\rm{AA}} \rangle =-\langle \chi_{\rm{BB}} \rangle $ acquire finite expectation values whereas $\langle \chi_{\rm{AB}} \rangle$ and $\langle \chi_{\rm{BA}} \rangle$ remain pinned to zero. Our self-consistent calculations indicate that the critical coupling $J_c/K$ for the $\mathbf{q}=\mathbf{q}_{1}$ solution is below it's $\mathbf{q}=0$ counterpart for the entire range of twist angles (Fig.~\ref{Fig_twist}~(a)), indicating that the modulated hybridization is energetically favored. A finite-$\mathbf{q}$ hybridization connects states near inequivalent Dirac points in the moir\'e BZ and gaps the spectrum, as illustrated in Fig.~\ref{Fig_twist}~(b), effectively `untwisting' the system. In contrast, for $\mathbf{q}=0$, the spectrum remains gapless, (Fig.~\ref{Fig_twist}~(c)). Consequently, the finite-$\mathbf{q}$ solution is preferred for any non-zero twist angle. The two solutions merge smoothly as $\theta \rightarrow 0$ since $\mathbf{q}_1$ vanishes in this limit, at which point the low-energy sectors match the self-consistent solutions of the un-twisted bilayer with AA stacking.  

At the level of the Hartree approximation, our results indicate that the gap remains open as the small-angle twisting is turned on. Within the same approximation, we conclude that resulting phases are adiabatically connected with the AA-stacked bilayer in the large-$J$ limit.
Our results suggest that, beyond the Hartree approximation, the gap in the spin excitations of the bilayer survives, and that the GS remains in a net zero-spin  state for small-angle twisting. We expect that the intra-layer interactions lift the extensive degeneracy of the orbitals, resulting in a gapped, $\mathbb{Z}_{2}$ QSL, as for the case with AA stacking.   

For the gauge choice of uniform and identical bonds in both layers, the $\mathbf{q}=\mathbf{q}_{1}$ incommensurate, inter-layer hybridization breaks the translation symmetry of simple moir\'e pattern but preserves all other symmetries. It consequently triples the size of the moir\'e unit cell (Fig. ~\ref{Fig_twist}~(d)). Fig.~\ref{Fig_twist}~(e) shows the moir\'e (black) and folded (red) BZ's, respectively. The rotated Dirac cones at the corners of the moir\'e BZ are folded onto the $\Gamma$ point. However, since the effective hybridization is not gauge invariant, this does not imply a true translation symmetry breaking, but instead demonstrates that small-angle twisting preserves the gapped $\mathbb{Z}_{2}$ QSL. 

We note that the main conclusion of the preceding paragraphs, that twisting the AA-stacked bilayer by small angles preserves the gapped spectrum, does not rely on our assumptions concerning the cutoff in $J(\mathbf{k})$. Indeed, keeping only the leading $J(0)$ terms in the expression for the self-consistent hybridization (SM eq. S82-S84), which likewise connect pairs of Dirac points in the moir\'e BZ, leads to a similar conclusion in the Hartree approximation.   

{\sl Discussion.} 
It is instructive to contrast the bilayer Yao-Lee model considered here with the bilayer Kitaev models of earlier works. For a bilayer Kitaev model, a mean-field study predicts gapped QSL and trivial dimer phases for intermediate and large values of the inter-layer coupling, respectively~\cite{Seifert_PRb2018}. However, an exact diagonalization study~\cite{Tomishige_PRB2019} finds that a single phase transition between gapless QSL and trivial dimer phases occurs at a substantially weaker coupling $J/K \sim 0.06$. Our results indicate that the QSL phase remains stable in Yao-Lee bilayers for large but finite intra-layer couplings, while the trivial dimer phase emerges only in the absence of intra-layer terms ($K=0$). The stability of the gapped QSL in the Yao-Lee bilayer can be attributed to the effect of the spin operators on the zero-flux GSs of the decoupled layers. In the Kitaev model, the spin operators create two visons, as shown in Sec. X of the SM. By contrast, the spin operators in the Yao-Lee model preserve the zero-flux GS manifold, since the spin and flux operators are associated with different DOF.       

{\sl Conclusion.} We studied the zero-temperature phase diagram of a bilayer Yao-Lee model with inter-layer interactions. For AA stacking, we determined that finite inter-layer singlet correlations gap the itinerant Majorana fermion spectrum. We also derived an effective Hamiltonian in the limit of large $J/K$, and demonstrated that it maps onto the toric code. In the absence of any additional transitions which close the gap, we concluded that the solutions obtained via the Hartree approximation are adiabatically connected to the large inter-layer interaction limit, leading to the stability of a topological gapped $\mathbb{Z}_{2}$ QSL. This phase persists for moir\'e superlattices under small-angle twisting. Detailed studies of the AB stacked phases and of the toric code models in the large inter-layer coupling limit are clearly desirable. 

We thank Piers Coleman and Filip Ronning for fruitful discussions. OE acknowledge support from NSF Award No.~DMR 1904716. MA is supported by Fulbright Scholarship.  This work was in part supported by the Deutsche Forschungsgemeinschaft under grants SFB 1143 (project-id 247310070) and the cluster of excellence ct.qmat (EXC 2147, project-id 390858490).

\clearpage
\pagebreak

\setcounter{equation}{0}
\setcounter{figure}{0}

\renewcommand{\theequation}{S\arabic{equation}}
\renewcommand{\thefigure}{S\arabic{figure}}
\renewcommand{\bibnumfmt}[1]{[S#1]}
\renewcommand{\citenumfont}[1]{S#1}

\widetext
\begin{center}
\textbf{Supplemental Materials for ``Kitaev spin-orbital bilayers and their moir\'e superlattices"}
\end{center}

\makeatletter

\section{Inter-layer singlets in the Majorana representation}

In the main text, we remark that a finite effective hybridization indicates the formation of static inter-layer singlet pairs. A similar connection has been discussed in previous works, which employed an exact Majorana representation of the spin operators~\cite{S_Shastry_PRB1997, S_Biswas_PRB_2011}. We summarize these arguments here. 

We introduce an equivalent basis of complex fermions by taking linear combinations of two itinerant Majorana fermions of the same flavor on overlapping sites in the two layers: 

\enni \begin{align}
f_{i}^{(\alpha)}=\frac{1}{2}\left(c_{1i}^{(\alpha)}+ic_{2i}^{(\alpha)} \right).
\end{align}

\enni The corresponding Fock space is determined by the three occupation numbers $n^{(\alpha)}_{i}= f^{\dag, (\alpha)}_{i} f^{(\alpha)}_{i}$, where $\alpha \in \{x, y, z\}$. 

To illustrate the connection between the Fock states and the Hilbert space of the local spins, we consider the inter-layer coupling for a single pair of overlapping sites:

\enni \begin{align}
H_{\text{I}, j}= & J \pmb{\sigma}_{1j}\cdot \pmb{\sigma}_{2j} 
\end{align}

\enni For antiferromagnetic interactions, this has a inter-layer singlet ground state and three excited triplet states at $4J$. When expressed in terms of the complex fermions, $H_{\text{I}, j}$ becomes

\enni \begin{align}
H_{\text{I}, j} = &-2J \left(\sum_{\alpha} n^{(\alpha)}_j-\frac{3}{2}\right)^2
\end{align}

\enni For $J>0$, there are two degenerate ground-state configurations with $n^{(x)}_{j}= n^{(y)}_{j}= n^{(z)}_{j}$ equal to 0 and 1, respectively, and six degenerate excited states at $4J$ for the remaining configurations. The Fock states thus provide two redundant representations of the product space of the two spins, which can be distinguished by the fermion parity $\prod_{\alpha} (2n^{(\alpha)}_{j}-1)$. The connection can also be made explicit by matching the matrix elements of the spin operators in either basis~\cite{S_Shastry_PRB1997, S_Biswas_PRB_2011},

In the complex fermion representation, the effective hybridization becomes

\enni \begin{align}
\langle \chi^{(\alpha)}_j \rangle = & \langle ic_{1j}^{(\alpha)} c_{2j}^{(\alpha)}\rangle 
\notag \\
= & \left( 2 n^{(\alpha)}_{j} - 1\right).
\label{Eq:lclprty}
\end{align}

\enni For the degenerate spin-singlet ground-state sector of even and odd fermion parities, $\braket{\chi^{(x)}_{j}}= \braket{\chi^{(y)}_{j}}= \braket{\chi^{(z)}_{j}}= \pm 1$, respectively. This statement can be generalized beyond a single pair of spins. For $J>0$, a local hybridization which is non-zero and equal for all three flavors indicates the presence of inter-layer singlets in the ground-state. 

In our calculations, the redundancy of the Majorana or complex fermion representations for pairs of spins was explicitly removed by choosing a gauge where all of the bond variables are equal to 1. 

\section{Projection onto physical space}

In the main text, we state that the expectation value of a gauge-invariant correlator in a state obtained by projecting our fixed-gauge ansatz onto the physical sector is consistent with a finite effective hybridization, for AA stacking. Here, we show that this is the case. 

\subsection{Effect of projection operator}

\label{Sec:Prjc}

We denote our ground-state ansatz with a set of finite $\braket{\chi_{i}}$ in a fixed gauge with uniform bonds equal to 1 by  

\enni \begin{align}
\ket{\Psi} = \ket{\forall~u_{1, ij}=1} \otimes \ket{\forall~u_{2, ij}=1} \otimes \ket{\braket{\chi_{\rm{AA}}}= - \braket{\chi_{\rm{BB}}}}.
\end{align}

\enni for $i, j \in \{1, 2, \hdots N \}$, where $N$ is the number of sites. We chose periodic boundary conditions for both layers and assumed an even number of unit cells $N_{c}=N/2$. In the trivially dimerized limit for $J \neq 0, K=0$ $\ket{\Psi}$ can be labeled by the eigenvalues of all $\chi_{i}$ operators. In this case, states obtained by flipping at least one of the $\braket{\chi_{i}}$'s are orthogonal to $\ket{\Psi}$. In the following, we assume that this can be generalized to ansatze where the $\chi_{i}$'s are not individually conserved for $K \neq 0$. 

We consider the operators 

\enni \begin{align}
D_{\nu,i}=-i c_{\nu,i}^{(x)} c_{\nu,i}^{(y)} c_{\nu,i}^{(z)} d_{\nu,i}^{(x)} d_{\nu,i}^{(y)} d_{\nu,i}^{(z)} ,
\end{align}

\enni where $\nu$ is a layer index, acting on $\ket{\Psi}$. $D_{\nu,i}$ anti-commutes with the bond operators $u^{(\alpha)}_{\nu, ij}=-id_{\nu,i}^{(\alpha)}d_{\nu,j}^{(\alpha)}$ in layer $\nu$ and with $\chi^{(\alpha)}_{i}$ for all three $\alpha$ flavors. Its effect on $\ket{\Psi}$ amounts to a $\mathbb{Z}_{2}$ gauge transformation which flips the three bonds emanating from site $i$ in layer $\nu$ and $\braket{\chi_{i}}$ for all flavors on the same site. Any two $D$ operators commute and obey $D^{2}_{\nu, i} =1$.

The projection operator $P$ is given by 

\enni \begin{align}
P= & 2^{-2N}\left[ \prod^{N}_{i=1} \left(\frac{1+ D_{1, i}}{2} \right) \right] \left[ \prod^{N}_{i=1} \left(\frac{1+ D_{2,i}}{2} \right) \right].
\end{align}

\enni For each layer, the product of $D_{\nu, i}$ operators over any subset of sites $\Lambda$ differs from that over the complementary set by the product over all sites:

\enni \begin{align}
\prod_{i \in \Lambda} D_{\nu, i} = \left( \prod_{j \notin \Lambda} D_{\nu, j} \right) \left( \prod^{N}_{k=1} D_{\nu, k} \right). 
\end{align}

\enni  We can also express the projection operator as~\cite{S_Pedrocchi_PRB_2011}

\enni \begin{align}
P= & 2^{-2N} \prod_{\nu} \left[ \left( \sum^{'}_{\{i\}} \prod_{i \in \{i \}} D_{\nu, i} \right)
\left( 1+ \prod^{N}_{i=1} D_{\nu, i} \right) \right]
\end{align}

\enni where the primed summations, involving products of at most $N/2$ operators, cover half of all possible combinations, and thus include $2^{N-1}$ separate realizations. The terms in the second parentheses can be expressed as 

\enni \begin{align}
\left( 1+ \prod^{N}_{i=1} D_{1, i} \right)
\left( 1+ \prod^{N}_{j=1} D_{2, j} \right) 
= & 2 \left( 1 + \prod^{N}_{i=1} D_{1, i} \right)\left( \frac{1 + P_{0}}{2} \right)
\end{align}

\enni where

\enni \begin{align}
P_{0}= \prod^{N}_{i=1} D_{1, i}  D_{2, i}
\end{align}

\enni  since 

\enni \begin{align}
\prod^{N}_{i=1} D_{2,i} = & \prod^{N}_{i=1} D_{1, i} P_{0}.
\end{align}

\enni The projection operator can therefore be written as 

\enni \begin{align}
P= & 2^{-2N+1}  \left( \sum^{'}_{\{i\}} \prod_{i \in \{i \}} D_{1, i} \right)
\left( \sum^{'}_{\{j\}} \prod_{j \in \{j \}} D_{2,j} \right)
 \left( 1 + \prod^{N}_{i=1} D_{1, i} \right)\left( \frac{1 + P_{0}}{2} \right)
\end{align}

\enni The effect of $P_{0}$ acting on $\ket{\Psi}$ is discussed further below. Of the remaining terms, the first parenthesis denotes a sum over all products of at most $N/2$ operators, each of which flips bond operators in layer 1 and the corresponding $\braket{\chi_{i}}$. The terms in the second parenthesis do the same in layer 2. The non-trivial part of the third parenthesis leaves all of the bond variables in both layers invariant, but flips all $\braket{\chi_{i}}$. The resulting non-trivial states differ from $\ket{\Psi}$ by at least three bonds in either layer or by a finite set of $\braket{\chi_{i}}$, and are therefore orthogonal by assumption. 

We now consider $P_{0}$, which can be re-cast as

\enni \begin{align}
P_{0}= & \prod^{N_{c}}_{l=1 } D_{1, l\rm{A}}  D_{1, l\rm{B}} D_{2, l\rm{A}} D_{2, l\rm{B}} 
\notag \\
= & (-1)^{N_{c}} \left[ \prod_{l} \prod_{\alpha} (i d^{1\alpha}_{l\rm{A}} d^{1\alpha}_{l\rm{B}})  \right]
\left[ \prod_{l} \prod_{\alpha} (i d^{2\alpha}_{l\rm{A}} d^{2\alpha}_{l\rm{B}})  \right]
\left[ \prod_{l} \prod_{\alpha} \left(2n^{(\alpha)}_{f,l \rm{A}} - 1 \right)  \left(2n^{(\alpha)}_{f,l\rm{B}} - 1 \right) \right]
\label{Eq:Prjc}
\end{align}

\enni where A and B denote the two sublattices and where we used the relation between $\chi_{i}$ and the local complex fermion parity introduced in Eq.~\ref{Eq:lclprty}. The $l$ indices label the $N_{c}$ unit cells. The terms in the first two brackets correspond to the total fermion parities of the $\mathbb{Z}_{2}$ gauge fields on layers 1 and 2, respectively~\cite{S_Chulliparambil_PRB2020}. The remaining terms determine the total parity of the itinerant fermions:

\enni \begin{align}
\left[ \prod_{l} \prod_{\alpha} \left(2n^{(\alpha)}_{f,l \rm{A}} - 1 \right)  \left(2n^{(\alpha)}_{f,l\rm{B}} - 1 \right) \right] = & (-1)^{\sum_{l} \sum_{\alpha} \left(n^{(\alpha)}_{f, l \rm{A}} + n^{(\alpha)}_{f, l \rm{B}} \right)}.
\label{Eq:Prty}
\end{align}

\enni The itinerant fermion parity thus depends on the total filling. 

As mentioned previously, we consider periodic boundary conditions along with an even number of unit cells along each of the two directions of the Bravais lattice. The fermion parities of the $\mathbb{Z}_{2}$ gauge fields are then both even~\cite{S_Chulliparambil_PRB2020}. Since the itinerant (complex) fermion sector is in a charge density wave phase at half filling, the fermion parity associated with these states is simply $(-1)^{3N_{c}}$. Consequently, the effect of $P_{0}$ on $\ket{\Psi}$ is trivial 

\enni  \begin{align}
\left( \frac{1 + P_{0}}{2} \right) \ket{\Psi} = \ket{\Psi}.
\end{align}

\enni This implies that $P$ acting on $\ket{\Psi}$ is 

\enni \begin{align}
P \ket{ \Psi}= & 2^{-2N+1}  \left( \sum^{'}_{\{i\}} \prod_{i \in \{i \}} D_{1, i} \right)
\left( \sum^{'}_{\{j\}} \prod_{j \in \{j \}} D_{2, j} \right)
 \left( 1 + \prod_{i} D_{1, i} \right) \ket{\Psi}.
 \label{Eq:PhGS}
\end{align}

\enni Note that the resulting state involves a linear combination over $2^{2N-1}$ distinct configurations.   

\subsection{Gauge-invariant correlator in the Majorana representation}

We can define a gauge-invariant operator~\cite{S_Fradkin_PRD_1979, S_Tsvelik_arxiv2021}

\enni \begin{align}
C^{(\alpha)}_{ij} = & \left( \prod^{'}_{\alpha-\text{links},\braket{i'j'}} u^{(\alpha)}_{1, i'j'} \right) \left( \prod^{''}_{\alpha-\text{links},\braket{i''j''}} u^{(\alpha)}_{2, i''j''} \right) \chi^{(\alpha)}_{i \rm{A}} \chi^{(\alpha)}_{j \rm{B}}    
\label{Eq:GICr}
\end{align}

\enni where $\prod^{'}$ and $\prod^{''}$ denote products of $u$ bonds in the upper and lower layer, respectively, which connect the two sites $i,j$ on A and B sublattices, respectively. For convenience, we choose overlapping paths in both layers. For the case of preserved SO(3) symmetry considered here, we drop the flavor indices. The expectation value of $C_{ij}$ is the same in any state $\ket{\Psi'}$ which is gauge equivalent to $\ket{\Psi}$. This is because $D_{\nu, k}$ either flips two bonds in the product of $u$'s in layer $\nu$ for $k \neq i, j$, or flips one bond and inverts $\braket{\chi_{i}}$ for $k \in \{i, j\}$, respectively. This operator is also invariant under gauge transformations which do not change any of the bonds but which flip all $\braket{\chi_{i}}$. $C_{ij}$ also preserves all of the bond variables. Therefore, we can write 

\enni \begin{align}
\braket{\Psi'|C_{ij}|\Psi'} = & \braket{\Psi|\chi_{i \rm{A}} \chi_{j \rm{B}}| \Psi}
\end{align}

\enni together with 

\enni \begin{align}
\lim_{ \left| \mathbf{R}_{i} - \mathbf{R}_{j} \right| \rightarrow \infty} \braket{\Psi'|C_{ij}|\Psi'} \approx \braket{\Psi| \chi_{\rm{A}} | \Psi} \braket{\Psi |\chi_{\rm{B}}| \Psi}.
\end{align}

\enni as the itinerant Majorana spectrum is gapped in this case. A finite $C_{ij}$ in any gauge and in the limit of asymptotically large separation is equivalent to a non-zero order parameter in our choice of gauge. 

We now determine the effect of the projector on the expectation values of $C_{ij}$. As shown previously, $P$ acting on $\ket{\Psi}$ generates a linear combination involving $2^{2N-1}$ distinct states. This implies that 

\enni \begin{align}
\braket{\Psi| P^{2}| \Psi} = & \left(2^{-4N+2} \right) \left( 2^{2N-1} \right)
\notag \\
= & 2^{-2N+1}
\end{align}

\enni where the two terms on the first line are due to the overall powers of $1/2$ and to the number of distinct configurations, respectively. Since $C_{ij}$ does not connect any two distinct configurations we obtain

\enni \begin{align}
\frac{\braket{\Psi|P C_{ij} P| \Psi}}{\braket{\Psi|P^{2} | \Psi}} = & \frac{ 2^{-2N+1} \braket{\Psi| C_{ij} | \Psi}} {2^{-2N+1}}
\notag \\
= & \braket{\Psi|\chi_{i \rm{A}} \chi_{j \rm{B}}| \Psi}
\end{align}

\enni Therefore, the expectation value of the gauge-invariant correlator in a projected ground-state ansatz is consistent with the the effective hybridization obtained in a fixed gauge without any additional projection. Our conclusions survive the projection to the physical space. 

\subsection{Gauge-invariant correlator in the spin and orbital basis}

In this section, we express the gauge-invariant correlator

\enni \begin{align}
C^{(\alpha)}_{ij} = & \left( \prod^{'}_{\alpha-\text{links},\braket{i'j'}} u^{(\alpha)}_{1, i'j'} u^{(\alpha)}_{2, i'j'} \right) \chi^{(\alpha)}_{i \rm{A}} \chi^{(\alpha)}_{j \rm{B}}    
\end{align}

\enni defined in the previous subsection in terms of the spin ($\sigma^{(\alpha)}_{\nu, i}$) and orbital ($\tau^{(\alpha)}_{\nu, i}$) operators of the Yao-Lee bilayer. 

We are interested in the expectation value of $C^{(\alpha)}_{ij}$ in the physical GS $P \Psi$ (Eq.~\ref{Eq:PhGS}), which is subject to the constraint

\enni \begin{align}
D_{\nu, i} P \Psi = & P \ket{\Psi}.
\end{align}

\enni We can thus identify $D_{\nu, i}=1$ to obtain

\enni \begin{align}
\tau^{(\alpha)}_{\nu,i} D_{\nu, i} =& \tau^{(\alpha)}_{\nu,i} 
\notag \\
= & -d^{(\alpha)}_{\nu, i} c^{(x)}_{\nu, i} c^{(y)}_{\nu, i} c^{(z)}_{\nu, i}. 
\end{align}

\enni Using this relation, together with the definition of the bond operators 

\enni \begin{align}
u^{(\alpha)}_{\nu, ij}=-id_{\nu,i}^{(\alpha)}d_{\nu,j}^{(\alpha)}
\end{align}

\enni we express the product of overlapping bonds on layers 1 and 2 as

\enni \begin{align}
u^{(\alpha)}_{1, i'j'} u^{(\alpha)}_{2, i'j'} = - \left( \tau^{(\alpha)}_{1, i'} \tau^{(\alpha)}_{1, j'} \tau^{(\alpha)}_{2, i'} \tau^{(\alpha)}_{2, j'} 
\right) 
\left( 
\chi^{(x)}_{i'} \chi^{(y)}_{i'} \chi^{(z)}_{i'}
\right)
\left( 
\chi^{(x)}_{j'} \chi^{(y)}_{j'} \chi^{(z)}_{j'}
\right).
\end{align}

\enni By substituting this expression into the expectation value of $C^{(\alpha)}_{ij}$ in the physical GS, we write

\enni \begin{align}
\braket{C^{(\alpha)}_{ij}}_{\text{Phys}} = 
\left \langle
- 
\prod^{'}_{\alpha-\text{links},\braket{i'j'}}
\left( \tau^{(\alpha)}_{1, i'} \tau^{(\alpha)}_{1, j'} \tau^{(\alpha)}_{2, i'} \tau^{(\alpha)}_{2, j'} 
\right) 
\left( 
\chi^{(x)}_{i'} \chi^{(y)}_{i'} \chi^{(z)}_{i'}
\right)
\left( 
\chi^{(x)}_{j'} \chi^{(y)}_{j'} \chi^{(z)}_{j'}
\right) \chi^{(\alpha)}_{i \rm{A}} \chi^{(\alpha)}_{j \rm{B}}  
\right \rangle_{\text{Phys}},
\end{align}

\enni where the overall minus sign is due to the odd number of bonds connecting sites $i,j$ on sublattices A and B, respectively.  
We make use of the following operator identities

\enni \begin{align}
\left(\chi^{(\alpha)}_i \right)^{2} = & 1 \\
\left( 
\chi^{(x)}_i \chi^{(y)}_i \chi^{(z)}_i
\right)
\chi^{(\alpha)}_{i \rm{A}}
= & - \sigma^{(\alpha)}_{1, i}
\sigma^{(\alpha)}_{2, i}
\end{align}

\enni and similarly for B, to determine the correlator as

\enni \begin{align}
\braket{C^{(\alpha)}_{ij}}_{\text{Phys}} = &
\left \langle
- 
\left( 
\sigma^{(\alpha)}_{1, i \rm{A}}
\sigma^{(\alpha)}_{2, i \rm{A}}
\right)
\left( \prod^{'}_{\alpha-\text{links}}
 \tau^{(\alpha)}_{1, i'} \tau^{(\alpha)}_{1, j'} \tau^{(\alpha)}_{2, i'} \tau^{(\alpha)}_{2, j'} 
\right) 
\left( 
\sigma^{(\alpha)}_{1, j \rm{B}}
\sigma^{(\alpha)}_{2, j \rm{B}}
\right) 
\right \rangle_{\text{Phys}},
\\
= & \left \langle
- 
\left( 
\sigma^{(\alpha)}_{1, i \rm{A}}
\sigma^{(\alpha)}_{2, i \rm{A}}
\right)
\left( \prod^{'}_{\alpha-\text{links}}
 p^{(\alpha)}_{i'} 
 p^{(\alpha)}_{j'}  
\right) 
\left( 
\sigma^{(\alpha)}_{1, j \rm{B}}
\sigma^{(\alpha)}_{2, j \rm{A}}
\right) 
\right \rangle_{\text{Phys}}
\label{Eq:PhCr}.
\end{align}

\enni We also used 

\enni \begin{align}
p_{i'}^{(\alpha)} = \tau_{1, i'}^{(\alpha)}\tau_{2, i'}^{(\alpha)},
\end{align}

\enni in the last expression.  

We  comment on the interpretation of this expectation value. As determined by the perturbative analysis in the large-$J$ limit,  presented in the main text, the physical GS manifold at zeroth order in the inter-layer interactions features decoupled, spin-singlets on overlapping sites  and free orbital degrees-of-freedom. The orbital states take on all values of $p^{(\alpha)}_{i'} 
 p^{(\alpha)}_{j'} = \pm 1$ for NN  $i',j'$. The average of the string operators over the GS manifold is zero , and $\braket{C^{(\alpha)}_{ij}}$ vanishes. By contrast, at second order ($g_{2}$) and beyond, we find that the GS manifold is constrained such that  $p^{(\alpha)}_{i'} 
 p^{(\alpha)}_{j'} = 1$. The   correlator acquires a finite expectation value, which is determined exclusively by the presence of the spin-singlets in the GS manifold. This indicates the emergence of a gapped, $\mathbb{Z}_{2}$ quantum spin liquid phase, as discussed in the main text.

\section{Variational analysis for finite flux configurations}
\begin{figure}[h!]
\includegraphics[width=0.5\columnwidth]{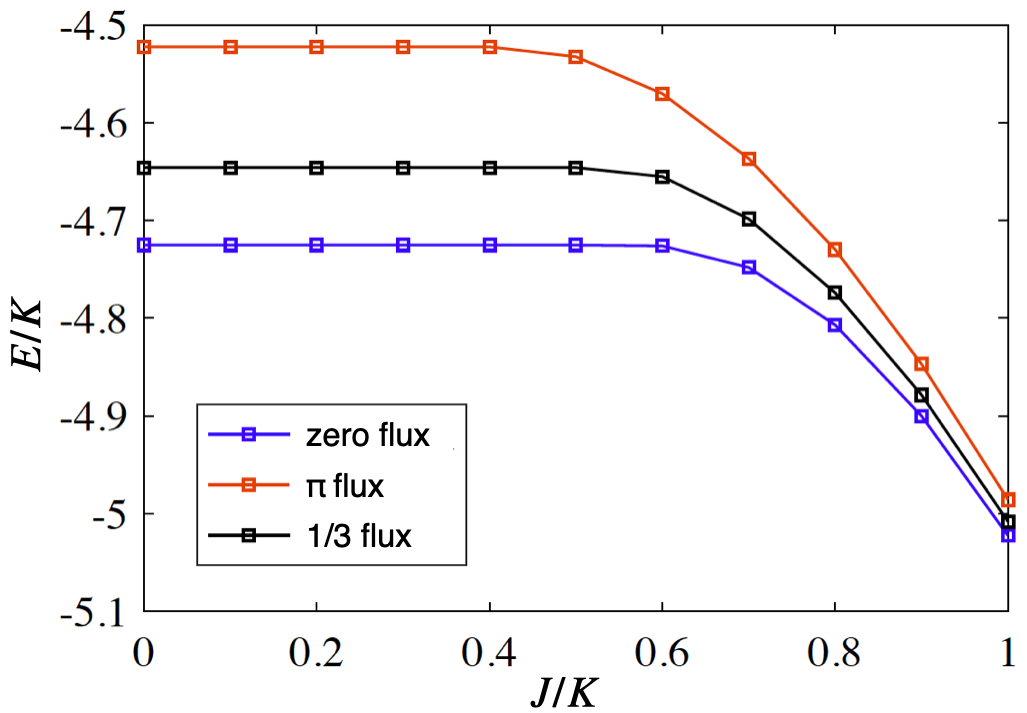}
\caption{Ground-state energy versus interlayer exchange $J$, 
for zero-, 1/3- and $\pi$-flux configurations. We find that the zero-flux configuration is the lowest in energy throughout.
}
\label{Fig:J_vs_E}
\end{figure}

In order to determine if the ground-state stays in the zero flux sector as a function of interlayer exchange $J$, we consider two additional flux configurations introduced in Ref. \citenum{S_Chulliparambil_PRB2021} in the context of the simplest Kitaev model an external magnetic field~\cite{S_Chulliparambil_PRB2021}. As shown in Fig. \ref{Fig:J_vs_E}, we find that zero flux configuration continues to be the lowest energy configuration among these three variational configurations.
 
\section{Derivation of the Effective Hamiltonian}

In this section, we provide the details of the derivation of the effective Hamiltonian in the large inter-layer exchange limit. We start with two singlets formed between layer 1 and 2 on sites $i$ and $j$, that are connected via an $\alpha$ ($\alpha = x, y,z$) bond. The unperturbed states are $|\phi \rangle = |S, \tau \rangle_i |S, \tau \rangle_j$ where $|S\rangle = |\uparrow_1 \downarrow_2 - \downarrow_1 \uparrow_2 \rangle$ is the singlet state and 1 and 2 are the layer subindices. $|\tau \rangle$ is the orbital component of the wave function. For $K=0$, the orbital sector is degenerate as there is no term in the Hamiltonian that couples to the $\tau$'s. We perturb the degenerate manifold of $|\phi\rangle$ with the $K$ term that couples the two singlets
\begin{eqnarray}
H_K =K\big[ \tau_{1i}^\alpha \tau_{1j}^\alpha (\sigma_{1i} \cdot \sigma_{1j})+ \tau_{2i}^\alpha \tau_{2j}^\alpha (\sigma_{2i} \cdot \sigma_{2j}) \big]
\end{eqnarray}
The first order correction to the energy vanishes $E^{(1)} = \langle \phi | H_K | \phi \rangle = 0$. The second order correction to the energy is 
\begin{eqnarray}
E^{(2)} &=& \frac{\sum_{m\ne \phi} |\langle m| H_K|\phi \rangle|^2} {E_{\phi}-E_m} \\ \nonumber
&=& -\frac{K^2}{2J}, ~~{\rm for~ \langle \tau_{1i}^{\alpha} \tau_{2i}^{\alpha}\tau_{1j}^{\alpha}\tau_{2j}^{\alpha} \rangle = 1}\\ \nonumber
&=& 0 , ~~{\rm for~ \langle \tau_{1i}^{\alpha} \tau_{2i}^{\alpha}\tau_{1j}^{\alpha}\tau_{2j}^{\alpha} \rangle = -1} \nonumber
\end{eqnarray}
where $ \langle \tau_{1i}^{\alpha} \tau_{2i}^{\alpha}\tau_{1j}^{\alpha}\tau_{2j}^{\alpha} \rangle $ is the eigenvalue of the inter-layer plaquette operator evaluated in the degenerate $|\phi\rangle$ manifold. This leads to the second order term in the effective Hamiltonian, $g_2 \tau_{1i}^{\alpha} \tau_{2i}^{\alpha}\tau_{1j}^{\alpha}\tau_{2j}^{\alpha}$ where $g_2 =-K^2/4J$.
The third order correction to the energy is 
\begin{eqnarray}
E^{(3)} &=& \sum_{m\neq \phi, n\neq \phi} \frac{\langle \phi|H_K|m\rangle \langle m| H_K | n\rangle \langle n |H_K |m\rangle)}{(E_{\phi}-E_m)(E_{\phi}-E_n)} \\ \nonumber
&=& -\frac{K^3}{J^2}(\langle \tau_{1i}^{\alpha} \tau_{1j}^{\alpha}\rangle +\langle \tau_{2i}^{\alpha} \tau_{2j}^{\alpha}\rangle)
\end{eqnarray}
which gives rise to the $g_3 (\tau_{1i}^{\alpha} \tau_{1j}^{\alpha} +\tau_{2i}^{\alpha} \tau_{2j}^\alpha)$ term with $g_3 = -K^3/J^2$. Apart from the pairwise interactions, we also consider a ring-exchange term around a honeycomb. The unperturbed states are the six singlet states with degenerate orbital wave functions: $|\phi \rangle = |S, \tau \rangle_i |S, \tau \rangle_j |S, \tau \rangle_k |S, \tau \rangle_l |S, \tau \rangle_m |S, \tau \rangle_n$ (see Fig. 2(d) in the main text). The sixth order correction to the energy that involves the ring exchange gives 
\begin{eqnarray}
E^{(6)} &=& +\frac{K^6}{J^5} (\langle \tau_{1i}^{x}\tau_{1j}^x \tau_{1j}^{z}\tau_{1k}^z \tau_{1k}^{y}\tau_{1l}^y \tau_{1l}^{x}\tau_{1m}^x \tau_{1m}^{z}\tau_{1n}^z \tau_{1n}^{y}\tau_{1i}^y + \tau_{2i}^{x}\tau_{2j}^x \tau_{2j}^{z}\tau_{2k}^z \tau_{2k}^{y}\tau_{2l}^y \tau_{2l}^{x}\tau_{2m}^x \tau_{2m}^{z}\tau_{2n}^z \tau_{2n}^{y}\tau_{2i}^y\rangle \\
& = & -\frac{K^6}{(8J)^5}( \langle W_p^{1}+W_p^{2} \rangle)
\end{eqnarray}
where $W_p^{1(2)} =\tau_{1(2)i}^z \tau_{1(2)j}^y \tau_{1(2)k}^x \tau_{1(2)l}^z \tau_{1(2)m}^y \tau_{1(2)n}^y$ is the flux operator for layer 1(2). Therefore the ring-exchange term is $g_6 (W_p^1 + W_p^2)$ with $g_6 = -K^6/(8J)^5$.

\section{Projecting the third order and sixth order terms onto the ground-state manifold}
As discussed in the main text, the eigenstates of the $g_2$ terms in the effective Hamiltonian are given in terms of the states $|p^x, p^y, p^z\rangle$ where $p^\alpha =\pm$ is the eigenvalue of the $p^\alpha = \tau_1^\alpha \tau_2^\alpha$ operator. The ground-state of $g_2$ term also need to satisfy the Ising Gauss law: $G_i^P = \prod_{\Ydown} r_{ij} = -1 $ where $r_{ij} = \pm 1$ for pairs of $(+,+)$ and $(-,-)$ eigenvalues of $p^{(\alpha)}_{i/j}$, respectively. However $|p^x, p^y, p^z\rangle$ states are not eigenstates of the $g_3$ and $g_6$ terms. Below, we present the matrix elements of $\tau_{1,2}^\alpha$ operators on the $|p^x, p^y, p^z\rangle$ states.
\begin{eqnarray}
\tau_1^{x} |-,-,-\rangle &= &- |-,+,+\rangle \\ \nonumber
\tau_2^{x} |-,-,-\rangle &= &+ |-,+,+\rangle \\ \nonumber
\tau_1^{y} |-,-,-\rangle &= &i |+,-,+\rangle \\ \nonumber
\tau_2^{y} |-,-,-\rangle &= &-i |+,-,+\rangle \\ \nonumber
\tau_1^{z} |-,-,-\rangle &= &+ |+,+,-\rangle \\ \nonumber
\tau_2^{z} |-,-,-\rangle &= &- |+,+,-\rangle \\ \nonumber
\tau_1^{x} |-,+,+\rangle &= &- |-,-,-\rangle \\ \nonumber
\tau_2^{x} |-,+,+\rangle &= &+ |-,-,-\rangle \\ \nonumber
\tau_1^{y} |-,+,+\rangle &= &i |+,+,-\rangle \\ \nonumber
\tau_2^{y} |-,+,+\rangle &= &i |+,+,-\rangle \\ \nonumber
\tau_1^{z} |-,+,+\rangle &= &+ |+,-,+\rangle \\ \nonumber
\tau_2^{z} |-,+,+\rangle &= &+ |+,-,+\rangle \\ \nonumber
\tau_1^{x} |+,-,+\rangle &= &+ |+,+,-\rangle \\ \nonumber
\tau_2^{x} |+,-,+\rangle &= &+ |+,+,-\rangle \\ \nonumber
\tau_1^{y} |+,-,+\rangle &= &-i |-,-,-\rangle \\ \nonumber
\tau_2^{y} |+,-,+\rangle &= &+i |-,-,-\rangle \\ \nonumber
\tau_1^{z} |+,-,+\rangle &= &+ |-,+,+\rangle \\ \nonumber
\tau_2^{z} |+,-,+\rangle &= &+ |-,+,+\rangle \\ \nonumber
\tau_1^{x} |+,+,-\rangle &= &+ |+,-,+\rangle \\ \nonumber
\tau_2^{x} |+,+,-\rangle &= &+ |+,-,+\rangle \\ \nonumber
\tau_1^{y} |+,+,-\rangle &= &-i |-,+,+\rangle \\ \nonumber
\tau_2^{y} |+,+,-\rangle &= &-i |-,+,+\rangle \\ \nonumber
\tau_1^{z} |+,+,-\rangle &= &+ |-,-,-\rangle \\ \nonumber
\tau_2^{z} |+,+,-\rangle &= &+ |-,-,-\rangle \\ \nonumber
\end{eqnarray}
We can summarize these matrix elements as follows: $\tau_{1(2)}^{x(y,z)}$ acting on $|p^x, p^y, p^z\rangle$ keeps the $x(y,z)$ eigenvalue the same while flipping the other two eigenvalues. Therefore the $g_6$ term acting on a plaquette flips the bond configuration, which gives rise to a term
\begin{eqnarray}
-\kappa \sum_{\hexagon} \big( | \hexagon \rangle \langle \bar{\hexagon}| + {\rm H.c.} \big)
\label{eq:kappa}
\end{eqnarray}
where $\hexagon$ and $\bar{\hexagon}$ are conjugate $p$ configurations around the hexagon. However, $g_3$ acting on a bond that obeys the Gauss's law breaks 4 bonds, which takes it outside the ground-state manifold. These virtual excitations can couple different ground-state configurations when $g_3$ term is applied around closed loops. The smallest loop is around a single honeycomb and when $g_3$ term applied around a honeycomb also lead to flipping the bond configuration as in eq. \ref{eq:kappa}. Since $\kappa\sim g_3^6/g_2^5$, it arises at $K^8/J^7$ order in perturbation theory.

\section{Mapping the ground-state manifold to toric code}
Kitaev's toric code~\cite{S_Kitaev_AnnPhys2003} is defined on a square lattice. However, it is straightforward to generalize it to a honeycomb lattice

\begin{eqnarray}
H_{TC} = - \kappa \sum_{\hexagon} W_{\hexagon} - \gamma_m \sum_{\Ydown} W_{\Ydown}
\end{eqnarray}
where $W_{\hexagon} = \prod_{\hexagon} \sigma^{z}_{ij}$ and $W_{\Ydown} = \prod_{\Ydown} \sigma^{x}_{ij}$.
The $\gamma_{m}$ terms define two sectors corresponding to states obeying even and odd Gauss’ laws, respectively. For $\gamma_{m}<0$ the odd sector is lowest in energy. The $\kappa$ terms amount to products of $\sigma^{z}_{ij}$ operators  which flip the $\sigma^{x}_{ij}$ bonds around each plaquette. These remove the extensive degeneracy of the odd Gauss' law sector and lead to a topological ground-state degeneracy instead. The same steps have been discussed in the effective model of the Yao-Lee bilayer in the large-$J$ limit. Hence, the extensive degeneracy of the ground-state manifold in the effective model is lifted in the same way, leading to an equivalent topological degeneracy.

\section{Effective hybridization for moir\'e superlattices}

In this section, we derive the mean-field Hamiltonian in the low-energy limit. For clarity, we shall use an expanded vector notation for the site indices. Our staring point is the interacting Hamiltonian  

\enni \begin{align}
\mathcal{H}= & \mathcal{H}_{1} + \mathcal{H}_{2} + \mathcal{H}_{\text{I}},
\end{align}

\enni where $\mathcal{H}_{1,2}$ consist of intra-layer terms for the respective layers, while $\mathcal{H}_{\text{I}}$ corresponds to the inter-layer interactions. Explicitly, these are

\enni \begin{align}
\mathcal{H}_{1} = &  \sum_{\mu} \sum_{\mathbf{R}} \sum_{\mathbf{l} } i K u_{1}(\mathbf{R}, \mathbf{l}) c^{(\mu)}_{A,1}(\mathbf{R}) 
c^{(\mu)}_{B,1} (\mathbf{R}+\mathbf{l}) 
\notag \\
\mathcal{H}_{2} = & \sum_{\mu} \sum_{\mathbf{R'}} \sum_{\mathbf{l'}}  i K u_{2}(\mathbf{R'}, \mathbf{l'}) c^{(\mu)}_{A,2}(\mathbf{R'}) 
c^{(\mu)}_{B,2} (\mathbf{R'}+\mathbf{l'}) 
\end{align}
 
\enni where $\mathbf{R},\mathbf{R}'$ are general Bravais lattice vectors of layer 1 and 2, respectively, while $\mathbf{l}, \mathbf{l}'$ are Bravais lattice vector corresponding to the three nearest-neighbor (NN) unit cells. In all subsequent sections, un-primed and primed vectors correspond to vectors in layers 1 and 2, respectively. Furthermore, all real-space vectors are determined w.r.t. the intersection of the twist axis with the respective planes. $\mu \in \{x, y, z \}$ stands for the flavor indices associated with both spin and orbital degrees-of-freedom (DOF). A and B are sublattice indices corresponding to 

\enni \begin{align}
c^{(\mu)}_{\rm{A},1}(\mathbf{R}) = & c^{(\mu)}_{1}(\mathbf{R}+ \pmb{\tau}_{\rm{A}}) 
\\
c^{(\mu)}_{\rm{B},1}(\mathbf{R}) = & c^{(\mu)}_{1}(\mathbf{R}+ \pmb{\tau}_{\rm{B}}) 
\end{align}  

\enni and similarly for layer 2. Note that the $\pmb{\tau}$'s depend on the stacking pattern. 

\enni \begin{align}
u_{1}(\mathbf{R}, \mathbf{l}) = & -i d^{(\mu)}_{A, 1}(\mathbf{R}) d^{(\mu)}_{B,1}(\mathbf{R}+ \mathbf{l})
\end{align}

\enni are the bond operators~\cite{S_Yao_PRL2011} consisting of two $d$ Majorana operators used in the representation of the local orbital DOF.  We use the same convention in defining the sublattice indices as for the itinerant $c$ Majorana operators. We choose a gauge where the bond operators are independent of the flavor indices, and consequently drop the latter from all subsequent expressions.  

The inter-layer spin-exchange interactions are

\enni \begin{align}
H_{\text{I}} = & \frac{1}{2} \sum_{\mu \neq \nu} \sum_{\alpha, \beta} \sum_{\mathbf{R},\mathbf{R'}} J(\mathbf{R}+\pmb{\tau}_{\alpha}-\mathbf{R'}-\pmb{\tau}'_{\beta}) c^{(\mu)}_{\alpha,1}(\mathbf{R}) c^{(\nu)}_{\alpha,1}(\mathbf{R}) c^{(\mu)}_{\beta,2}(\mathbf{R'})  c^{(\nu)}_{\beta,2}(\mathbf{R'}).
\end{align}

\enni The itinerant Majorana fermions obey 

\enni \begin{align}
c^{\dag, (\mu)}_{\alpha, 1}(\mathbf{R}) = & c^{ (\mu)}_{\alpha, 1}(\mathbf{R})
\end{align}

\enni together with

\enni \begin{align}
\{ c^{(\mu)}_{\alpha,1}(\mathbf{R}), c^{(\nu)}_{\beta,1}(\mathbf{R'}) \} = & 2 \delta_{\mathbf{R}, \mathbf{R'}} \delta_{\mu, \nu} \delta_{\alpha, \beta},
\end{align}

\enni and similarly for layer 2.  

We next consider the expansion of the $c$ Majorana fermions in terms of Bloch waves. Tilde momenta in the two layers are measured w.r.t. the intersections of the planes with the twist axis. Momenta without tilde are defined only in the vicinity of Dirac points in either layers, and are assumed to include a large number of moir\'e reciprocal unit cells for the small twist angles considered here.  Finally, un-primed momenta correspond to layer 1, while primed momenta denote the layer 2 counterparts. With these conventions, we write 

\enni \begin{align}
c^{(\mu)}_{\alpha,1}(\mathbf{R}) = & \frac{\sqrt{2}}{\sqrt{N}} \sum_{\mathbf{k} \in C/2} 
\left[ 
e^{-i \mathbf{\tilde{k}} \cdot (\mathbf{R} + 
\pmb{\tau}_{\alpha}) } c^{(\mu)}_{\alpha,1}(\tilde{\mathbf{k}}) 
+  e^{i \mathbf{\tilde{k}} \cdot (\mathbf{R} + 
\pmb{\tau}_{\alpha}) } c^{\dag, (\mu)}_{\alpha,1}(\tilde{\mathbf{k}}) 
\right]
\notag \\
c^{(\nu)}_{\alpha,2}(\mathbf{R'}) = & \frac{\sqrt{2}}{\sqrt{N}} \sum_{\tilde{\mathbf{k'}} \in C'/2} 
\left[ 
e^{-i \tilde{\mathbf{k'}} \cdot (\mathbf{R'} + 
\pmb{\tau}'_{\alpha}) } c^{(\nu)}_{\alpha,2}(\mathbf{\tilde{k'}}) 
+  e^{i \mathbf{\tilde{k'}} \cdot (\mathbf{R'} + 
\pmb{\tau}'_{\alpha}) } c^{\dag, (\nu)}_{\alpha,2}(\mathbf{\tilde{k'}}) 
\right],
\label{Eq:FT}
\end{align}
 
\enni where $\alpha, \beta$ are sublattice indices, and $N$ is the number of unit cells, assumed identical in either layer. The Majorana nature implies that~\cite{S_Chulliparambil_PRB2020}

\enni \begin{align}
c^{(\mu), \dag}_{\alpha, 1}(\tilde{\mathbf{k}}) = &  c^{(\mu)}_{\alpha, 1}(-\tilde{\mathbf{k}})
\end{align}

\begin{figure}[h!]
\includegraphics[width=0.5\columnwidth]{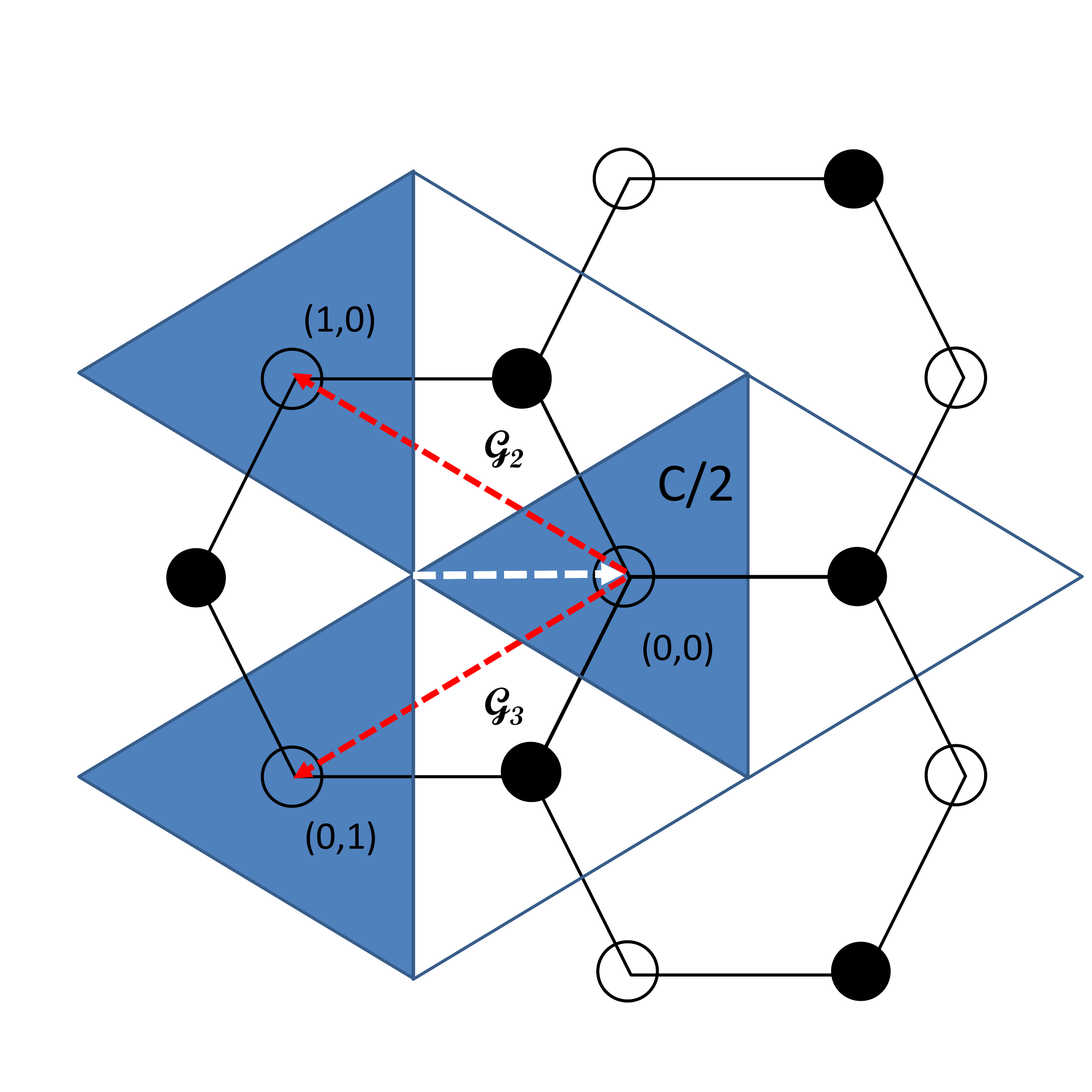}
\caption{Extended BZ for layer 1. The blue shaded areas represent our choice of (half) unit cells here, labeled by $C/2$. Note that these are equivalent to the unit cell shown in Fig.~3~(d) of the main text. The white vector indicates the position of the Dirac point in the first BZ w.r.t. to the origin, which coincides with the twist axis. The red vectors are the reciprocal unit vectors $\pmb{\mathcal{G}}_{2}$ and $\pmb{\mathcal{G}}_{3}$. The pairs of indices label the valleys. A twisted variant of this figure can be drawn for layer 2.}
\label{Fig:Vctr_dfnt}
\end{figure}

\enni where $\mathbf{G}$ is a reciprocal vector. This redundancy is accounted for in Eqs.~\ref{Eq:FT} by restricting the sums to one half of the primitive reciprocal unit cell, as shown in Fig.~\ref{Fig:Vctr_dfnt}. The operators defined on $C/2~(C'/2)$ obey the standard anti-commutation relations

\enni \begin{align}
\{c^{\dag, (\mu)}_{\alpha, 1}(\tilde{\mathbf{k}}_{1}), c^{(\nu)}_{\beta, 1}(\tilde{\mathbf{k}}_{2})  \}= & \delta_{\tilde{\mathbf{k}}_{1}, \tilde{\mathbf{k}}_{2}} \delta_{\mu, \nu} \delta_{\alpha, \beta}.
\end{align}

\enni We assume periodic boundary conditions consistent with the uniform gauge choice adopted throughout the remaining sections, which imply the Bloch periodicity 
\enni \begin{align}
e^{-i  \mathbf{G} \cdot \pmb{\tau}_{\alpha}} c^{\dag, (\mu)}_{\alpha,1}(\tilde{\mathbf{k}}) = &  c^{\dag, (\mu)}_{\alpha,1}(\tilde{\mathbf{k}} + \mathbf{G}),
\end{align}

\enni and similarly for layer 2, where $\mathbf{G}$ is a reciprocal lattice vector. In the following, it will prove convenient to extend the summations in Eq.~\ref{Eq:FT} to $N_{c}$ (half) reciprocal unit cells in the extended Brillouin Zone (BZ). Although such a procedure is redundant, it illustrates the emerging moir\'e periodicity in the low-energy limit. 

\subsection{Intra-layer terms in the low-energy limit}  

\label{Sec:Intra}

We work is a gauge where both $u_{1,2}$ are uniform and equal to 1. The intra-layer terms are

\enni \begin{align}
\mathcal{H}_{1} 
= & \frac{1}{N_{c}} \sum_{\mu} \sum_{\tilde{\mathbf{k}}} 
 2iK f_{1}(\tilde{\mathbf{k}})  c^{\dag, (\mu)}_{A,1}(\tilde{\mathbf{k}}) c^{(\mu)}_{B,1}(\tilde{\mathbf{k}})
+ \text{H.c.}
\\
\mathcal{H}_{2} 
= & \frac{1}{N_{c}} \sum_{\mu} \sum_{\tilde{\mathbf{k}}'} 
 2iK f_{2}(\tilde{\mathbf{k}}')  c^{\dag, (\mu)}_{A,2}(\tilde{\mathbf{k}}') c^{(\mu)}_{B,2}(\tilde{\mathbf{k}}')
+ \text{H.c.} 
\end{align} 

\enni where the momenta sums cover an extended BZ of half $N_{c}$ primitive cells. 

\enni \begin{align}
f_{1}(\tilde{\mathbf{k}}) = & \sum_{\mathbf{l}} e^{i\tilde{\mathbf{k}} \cdot(\pmb{\tau}_{A} - \pmb{\tau}_{B}-\mathbf{l})}
\\
f_{2}(\tilde{\mathbf{k}}') = & \sum_{\mathbf{l}'} e^{i\tilde{\mathbf{k}'} \cdot(\pmb{\tau}'_{A} - \pmb{\tau}'_{B}-\mathbf{l}')}
\end{align}

\enni are the familiar graphene form factors, with $\mathbf{l}$ as defined previously. Note that these obey

\enni \begin{align}
f_{1}(\tilde{\mathbf{k}} + \mathbf{G}) e^{-i\mathbf{G}\cdot (\pmb{\tau}_{A}-\pmb{\tau_{B}} )}
= & f(\tilde{\mathbf{k}}).
\end{align}

\enni and similarly for $f_{2}$.

We now proceed to take the low-energy limits of $\mathcal{H}_{1,2}$. Due to twisting, the Dirac points are shifted to  

\enni \begin{align}
\mathbf{K}_{nm}= & \mathbf{K}_{00} + n \pmb{\mathcal{G}}_{2} + m \pmb{\mathcal{G}}_{3} \\
\mathbf{K}'_{nm}= & \mathbf{K}'_{00} + n \pmb{\mathcal{G}}'_{2} + m \pmb{\mathcal{G}}'_{3}
\end{align}

\enni for layers 1 and 2, respectively. The $n,m$ indices label the reciprocal unit cell translated form the first BZ at $n=m=0$ by a reciprocal lattice vector 

\enni \begin{align}
\mathbf{G} = n \pmb{\mathcal{G}}_{2} + m \pmb{\mathcal{G}}_{3},
\end{align}

\enni and similarly for layer 2. We expand the functions $f_{1,2}$ for a common set of momenta 

\enni \begin{align}
\tilde{\mathbf{k}} = \mathbf{K}_{00} + n \pmb{\mathcal{G}}_{2} + m \pmb{\mathcal{G}}_{3} +
\mathbf{k}
\end{align}

\enni with $\mathbf{k}$ restricted to be in the vicinity of the Dirac points. Using the Bloch periodicity, we obtain

\enni \begin{align}
H_{1} 
= & \sum_{\mu} \sum_{\mathbf{k}} 
 2iK \left[ \mathbf{k} \cdot \pmb{\nabla} f_{1}(\mathbf{K}_{00}) \right]  c^{\dag, (\mu)}_{A,1}(\mathbf{K}_{00} + \mathbf{k}) c^{(\mu)}_{B,1}(\mathbf{K}_{00} + \mathbf{k})
+ \text{H.c.}
\end{align} 

\enni Since $\mathbf{k}$ covers a large number of reciprocal moir\'e primitive unit cells, we can trivially extend the expression above to include shifted Dirac points as

\enni \begin{align}
H_{1} 
= & \frac{1}{N_{c}} \sum_{\mu} \sum_{n,m} \sum_{\mathbf{k}} 
 2iK \left[ \left( \mathbf{k} - n \mathbf{b_{2}} - m\mathbf{b}_{3} \right) \cdot \mathbf{\nabla} f_{1}(\mathbf{K}_{00}) \right]  c^{\dag, (\mu)}_{A,1}(\mathbf{K}_{00} + \mathbf{k} - n \mathbf{b_{2}} - m\mathbf{b}_{3}) c^{(\mu)}_{B,1}(\mathbf{K}_{00} + \mathbf{k} - n \mathbf{b_{2}} - m\mathbf{b}_{3})
+ \text{H.c.}
\end{align} 
 
\enni where 

\enni \begin{align}
\mathbf{b}_{2} = & \pmb{\mathcal{G}}'_{2} - \pmb{\mathcal{G}}_{2} \label{Eq:Mr_rcpr_1}\\
\mathbf{b}_{3} = & \pmb{\mathcal{G}}'_{3} - \pmb{\mathcal{G}}_{3} \label{Eq:Mr_rcpr_2}
\end{align}

\enni are the moir\'e reciprocal unit vectors. We can be re-write $\mathcal{H}{1}$ in compact form as 

\enni \begin{align}
\mathcal{H}_{1} 
= & \frac{1}{N_{c}} \sum_{\mu} \sum_{n,m} \sum_{\mathbf{k}} 
 2iK \left[ \left( \mathbf{k} - n \mathbf{b_{2}} - m\mathbf{b}_{3} \right) \cdot \pmb{\nabla} f_{1}(\mathbf{K}_{00}) \right]  c^{\dag, (\mu)}_{A,1;nm}(\mathbf{k} ) c^{(\mu)}_{B,1;nm}(\mathbf{k})
+ \text{H.c.}
\end{align} 

\enni where we introduced valley indices as in

\enni \begin{align}
c^{\dag, (\mu)}_{A,1}(\mathbf{K}_{00} + \mathbf{k} - n \mathbf{b_{2}} - m\mathbf{b}_{3}) = c^{\dag, (\mu)}_{A,1; nm}(\mathbf{k}).
\end{align}

The same steps can be applied to the layer 2 terms, provided that we take into account the shift of the Dirac points w.r.t. those of layer 1, together with a rotation in the Fermi velocities due to the rotation of the Bravais lattice vectors entering the definition of $f_{2}$: 

\enni \begin{align}
\mathcal{H}_{2} 
= & \frac{1}{N_{c}} \sum_{\mu} \sum_{n,m} \sum_{\mathbf{k}} 
 2iK \left[ \left( \mathbf{k} - \mathbf{q}_{1}-n \mathbf{b_{2}} - m\mathbf{b}_{3} \right) \cdot \hat{R}(\theta) \pmb{\nabla} f_{1}(\mathbf{K}_{00}) \right]  c^{\dag, (\mu)}_{A,2;nm}(\mathbf{k} ) c^{(\mu)}_{B,2;nm}(\mathbf{k})
+ \text{H.c.}
\end{align}  

\enni where 

\enni \begin{align}
\mathbf{q}_{1} = & \mathbf{K}'_{00} - \mathbf{K}_{00}
\label{Eq:Shft}
\end{align}

\enni is the relative shift of the Dirac points of layer 2 and 1 in the first BZ. The matrix $\hat{R}(\theta)$ is an in-plane rotation by the total relative twist angle $\theta$. The valley indices for layer 2 are defined precisely as for layer 1. 

\subsection{Inter-layer interactions in the low-energy limit}

We first proceed to decouple the inter-layer interactions in the paramagnetic channel as 

\enni \begin{align}
\mathcal{H}_{\text{I}} = - & \frac{1}{2} \sum_{\mu \neq \nu} \sum_{\alpha, \beta} \sum_{\mathbf{R},\mathbf{R'}} J(\mathbf{R}+\pmb{\tau}_{\alpha}-\mathbf{R'}-\pmb{\tau}'_{\beta}) 
\braket{c^{(\mu)}_{\alpha,1}(\mathbf{R}) 
c^{(\mu)}_{\beta,2}(\mathbf{R'}) } 
c^{(\nu)}_{\alpha,1}(\mathbf{R})  c^{(\nu)}_{\beta,2}(\mathbf{R'}).
\end{align}

\enni We introduce the Bloch wave expansions of Eq.~\ref{Eq:FT} and carry out the sums over the Bravais lattice vectors. For our choice of half reciprocal unit cells $C/2$ and $C'/2$ (Fig.~\ref{Fig:Vctr_dfnt}), pairing terms at opposite momenta have vanishing weight and are ignored. For the remaining terms we use

\enni \begin{align}
\sum_{\mathbf{R}}  J(\mathbf{R}+ \pmb{\tau}_{\alpha} - \mathbf{R}' - \pmb{\tau}_{\beta}') e^{ikR} 
\approx & \sum_{\mathbf{G}} e^{-i(\mathbf{k}+\mathbf{G})(\pmb{\tau}_{\alpha} - \mathbf{R}' - \pmb{\tau}_{\beta}') } J(\mathbf{k}+\mathbf{G})
\end{align}
 
\enni via the Poisson summation formula, where

\enni \begin{align}
J(\mathbf{k}+\mathbf{G}) = \frac{1}{a} \int_{a} d^{2} r J(\mathbf{r}) e^{i(\mathbf{k}+ \mathbf{G}) \cdot \mathbf{r}}
\end{align}

\enni is the Fourier transform of $J$, defined over the unit cell of layer 1 with an area $a$. Also taking into account the conservation of momentum, we obtain 

\enni \begin{align}
\mathcal{H}_{\text{I}}= &  -\frac{1}{N^{2}_{c}} \sum_{\tilde{\mathbf{k}}, \tilde{\mathbf{s}}} \sum_{\mathbf{G}, \mathbf{G}'} \sum_{\mu} \sum_{\alpha, \beta}
 \bigg\{ t^{(\alpha \beta)}_{I} \left( \tilde{\mathbf{k}}+ \mathbf{G}, \tilde{\mathbf{s}} \right)      
 e^{-i \mathbf{G} \cdot \pmb{\tau}_{\alpha}} 
e^{i \mathbf{G}' \cdot \pmb{\tau}'_{\beta}}  c^{\dag, (\mu)}_{\alpha,1}(\tilde{\mathbf{k}}) 
c^{(\mu)}_{\beta,2}(\tilde{\mathbf{k}}+\mathbf{G} - \mathbf{G}' - \tilde{\mathbf{s}} ) 
\notag \\
+& t^{(\alpha \beta)}_{II} \left(\tilde{\mathbf{k}}+\mathbf{G}, \tilde{\mathbf{s}} \right)      
 e^{-i \mathbf{G} \cdot \pmb{\tau}_{\alpha}} 
e^{i \mathbf{G}' \cdot \pmb{\tau}'_{\beta}}  c^{\dag, (\mu)}_{\alpha,1}(\tilde{\mathbf{k}}) 
c^{(\mu)}_{\beta,2}(\tilde{\mathbf{k}}+\mathbf{G} - \mathbf{G}' + \tilde{\mathbf{s}} ) 
\bigg\}+  \text{H.c.},
\end{align}

\enni where

\enni \begin{align}
t^{(\alpha \beta)}_{I} \left( \tilde{\mathbf{k}}+ \mathbf{G}, \tilde{\mathbf{s}} \right)  = & \frac{4}{N N^{2}_{c}}
\sum_{\tilde{\mathbf{k}}_{a}} J\left( 
\tilde{\mathbf{k}} + \mathbf{G} - \tilde{\mathbf{k}}_{a}
\right)      
\braket{
 c^{(\nu)}_{\alpha,1}(\tilde{\mathbf{k}}_{a}) 
c^{\dag, (\nu)}_{\beta,2}(\tilde{\mathbf{k}}_{a}-\tilde{\mathbf{s}}) 
}
\notag \\
t^{(\alpha \beta)}_{II}\left( \tilde{\mathbf{k}}+ \mathbf{G}, \tilde{\mathbf{s}} \right)  = & \frac{4}{N N^{2}_{c}}
\sum_{\tilde{k}_{a}} J\left(\tilde{\mathbf{k}} + \mathbf{G} + \tilde{\mathbf{k}}_{a}\right)      
\braket{
 c^{\dag, (\nu)}_{\alpha,1}(\tilde{\mathbf{k}}_{a}) 
c^{(\nu)}_{\beta,2}(\tilde{\mathbf{k}}_{a}-\tilde{\mathbf{s}}) 
},
\end{align}

\enni where we accounted for a mean-field ansatz which preserves the spin SO(3) symmetry, implying expectation values which are independent of the  flavor indices. Note that the sums over tilde momenta cover the extended BZ. 

The form of $\mathcal{H}_{\text{I}}$ is strongly reminiscent of the inter-layer hybridization in twisted bilayer graphene~\cite{S_Bistritzer_PNAS2011}. We can show that it reduces to a sum over an extended moir\'e BZ in the low-energy limit by applying the steps of Sec.~\ref{Sec:Intra}:

 \enni \begin{align}
& \mathcal{H}_{\text{I}} \approx  -\frac{1}{N_{c}} \sum_{\mathbf{k}, \mathbf{s}} \sum_{n', m'} \sum_{n'',m''} \sum_{\mu} \sum_{\alpha \beta} \bigg\{
 t^{(\alpha \beta)}_{I; n' m', n'' m''} \left( \mathbf{k}, \mathbf{s} \right) 
e^{-i\left( (n''-n') \pmb{\mathcal{G}}_{2} + (m''-m')\pmb{\mathcal{G}}_{3}\right) \cdot( \pmb{\tau}_{\alpha} - \pmb{\tau}_{\beta})}
 c^{\dag, (\mu)}_{1, \alpha; n'm'}(\mathbf{k}) 
c^{(\mu)}_{2, \beta; n'' m''}(\mathbf{k} - \mathbf{s}) 
\notag \\
+ & t^{(\alpha \beta)}_{II; n' m', n'' m''}\left( \mathbf{k}, \mathbf{s} \right) 
e^{-i\left( (n''-n') \pmb{\mathcal{G}}_{2} + (m''-m')\pmb{\mathcal{G}}_{3}\right) \cdot( \pmb{\tau}_{\alpha} - \pmb{\tau}_{\beta})}
 c^{\dag, (\mu)}_{1, \alpha; n' m'}(\mathbf{k}) 
c^{(\mu)}_{2, \beta; n'' m''}(\mathbf{k} + \mathbf{s})
+ \text{H.c.} \bigg\}
\end{align}
 
\enni where

\enni \begin{align}
& t^{(\alpha \beta)}_{I; n' m', n'' m''} \left( \mathbf{k}, \mathbf{s} \right) 
\notag \\
= &  
\frac{4}{N_{c} N} \sum_{n_{a}, m_{a}}
\sum_{\mathbf{k}_{a}} J\left( (n''-n'-n_{a}) \pmb{\mathcal{G}}_{2} + (m''-m'-m_{a})\pmb{\mathcal{G}}_{3}\right) 
 e^{i\left(n_{a} \mathcal{G}_{2} + m_{a}\mathcal{G}_{3} \right) \cdot(\pmb{\tau}_{\alpha} -\pmb{\tau}_{\beta})} \braket{
 c^{(\nu)}_{\alpha,1; 00}(\mathbf{k}_{a}) 
c^{\dag, (\nu)}_{\beta,2; n_{a} m_{a}}(\mathbf{k}_{a}-\mathbf{s}) 
}
\label{Eq:t_I}
\end{align}

\enni \begin{align}
& t^{(\alpha \beta)}_{II; n' m', n'' m''}\left( \mathbf{k}, \mathbf{s} \right) 
\notag \\
= & \frac{4}{N_{c} N} \sum_{n_{a}, m_{a}}
\sum_{\mathbf{k}_{a}} J\left( 2\mathbf{K}_{00}+(n''-n'+n_{a}) \pmb{\mathcal{G}}_{2} + (m''-m'+m_{a})\pmb{\mathcal{G}}_{3}\right) 
 e^{-i\left(n_{a} \mathcal{G}_{2} + m_{a}\mathcal{G}_{3} \right) \cdot(\pmb{\tau}_{\alpha} -\pmb{\tau}_{\beta})}  \braket{
 c^{\dag, (\nu)}_{\alpha,1;00}(\mathbf{k}_{a}) 
c^{(\nu)}_{\beta,2;n_{a} m_{a}}(\mathbf{k}_{a} - \mathbf{s}) 
}
\label{Eq:t_II}
\end{align}

\enni These terms represent an effective hybridization between states on layer 1, with Dirac points periodically extended throughout the moir\'e BZ zone, and all states on layer 2, with Dirac points which are shifted by a fixed vector $\mathbf{q}_{1}$ (Eq.~\ref{Eq:Shft}). This expression is invariant up to a phase under a translation by moir\'e reciprocal vectors (Eqs.~\ref{Eq:Mr_rcpr_1},~\ref{Eq:Mr_rcpr_2}). Also note that all sums involve vectors in the vicinity of the pair of Dirac points in the first BZ.  In addition, we assumed that the Fourier transform of $J$ varies slowly on the scale of a single Moire reciprocal unit cell.    

We further simplify these expression via the following three assumptions. First, we restrict the intermediate summations over $n_{a}, m_{a}$ to the leading 7 terms corresponding to $J\left(0 \right), J\left( \pm \pmb{\mathcal{G}}_{2} \right), J\left( \pm \pmb{\mathcal{G}}_{3} \right),J\left( \pm \pmb{\mathcal{G}}_{2}  \mp \pmb{\mathcal{G}}_{3} \right)$ for $t_{I}$, and the leading 6 terms  corresponding to $ J\left(2 \mathbf{K}_{00}  \right)$, $J\left(2 \mathbf{K}_{00} + 2\pmb{\mathcal{G}}_{2} \right)$, $ J\left(2 \mathbf{K}_{00} + 2\pmb{\mathcal{G}}_{3} \right)$, $J\left(2 \mathbf{K}_{00} + \pmb{\mathcal{G}}_{2} \right)$, $J\left(2 \mathbf{K}_{00} + \pmb{\mathcal{G}}_{3} \right)$, $J\left(2 \mathbf{K}_{00} +  \pmb{\mathcal{G}}_{2} + \pmb{\mathcal{G}}_{3} \right)$ for $t_{II}$. These explicitly preserve a $C_{3}$ rotation symmetry. Secondly, we restrict the  hybridization to states corresponding to NN Dirac points in the extended moir\'e zone. For given $n', m'$, this is done by imposing 

\enni \begin{align}
n''\mathbf{b}_{2} + m''\mathbf{b}_{3} + \mathbf{s} = \mathbf{q} + n'\mathbf{b}_{2} + m'\mathbf{b}_{3} + 
\begin{cases}
0 \\
\mathbf{b}_{2} \\
\mathbf{b}_{3}
\end{cases}
\end{align}

\enni, with $\mathbf{q}$ restricted to lie inside a moir\'e reciprocal unit cell, and by subsequently eliminating the sums over $n'', m''$ and $\mathbf{s}$. States near neighboring Dirac points are expected to provide the leading contributions to the effective hybridization in the low-energy limit. 
For convenience, we include the phase factors in Eqs.~
\ref{Eq:t_I},~\ref{Eq:t_II} as

\enni \begin{align}
\tilde{t}^{(\alpha \beta)}_{I; n' m', n'' m''} \left( \mathbf{k}, \mathbf{s} \right) 
= & t^{(\alpha \beta)}_{I; n' m', n'' m''} \left( \mathbf{k}, \mathbf{s} \right) e^{-i\left( (n''-n') \pmb{\mathcal{G}}_{2} + (m''-m')\pmb{\mathcal{G}}_{3}\right)} 
\end{align}

\enni \begin{align}
\tilde{t}^{(\alpha \beta)}_{II; n' m', n'' m''} \left( \mathbf{k}, \mathbf{s} \right) 
= & t^{(\alpha \beta)}_{II; n' m', n'' m''} \left( \mathbf{k}, \mathbf{s} \right) e^{-i\left( (n''-n') \pmb{\mathcal{G}}_{2} + (m''-m')\pmb{\mathcal{G}}_{3}\right)} 
\end{align}

With these assumptions, the only allowed terms for fixed $\mathbf{q}$ are 

\enni \begin{align} 
\tilde{t}
^{(\alpha \beta)}_{I; n' m', n' m'} \left( \mathbf{k}, \mathbf{q} \right)
=  &
\frac{4}{N} 
\sum_{\mathbf{k}_{a}} 
\bigg\{ 
  J\left(0 \right) 
 \braket{
 c^{(\nu)}_{\alpha,1;00}(\mathbf{k}_{a}) 
c^{\dag, (\nu)}_{\beta,2;00}(\mathbf{k}_{a} -\mathbf{q}) 
} 
+ J\left( -\pmb{\mathcal{G}}_{2} \right)
 e^{i \pmb{\mathcal{G}}_{2}( \pmb{\tau}_{\alpha} - \pmb{\tau}_{\beta})} 
 \braket{
 c^{(\nu)}_{\alpha,1;00}(\mathbf{k}_{a}) 
c^{\dag, (\nu)}_{\beta,2;10}(\mathbf{k}_{a} -\mathbf{q}) 
} 
\notag \\
& + J\left( -\pmb{\mathcal{G}}_{3} \right) 
e^{i \pmb{\mathcal{G}}_{3}( \pmb{\tau}_{\alpha} - \pmb{\tau}_{\beta})} \braket{  
 c^{(\nu)}_{\alpha,1;00}(\mathbf{k}_{a}) 
c^{\dag, (\nu)}_{\beta,2;01}(\mathbf{k}_{a} -\mathbf{q})} 
\bigg\}
\end{align}

\enni \begin{align}
\tilde{t}^{(\alpha \beta)}_{I; n' m', n'+1 m'} \left( \mathbf{k}, \mathbf{q} \right)
  = & 
\frac{4}{N} 
\sum_{\mathbf{k}_{a}} 
\bigg\{ 
  J\left(0 \right) 
 \braket{
 c^{(\nu)}_{\alpha,1;00}(\mathbf{k}_{a}) 
c^{\dag, (\nu)}_{\beta,2;10}(\mathbf{k}_{a} -\mathbf{q}) 
} 
+ J\left( \pmb{\mathcal{G}}_{2} \right)
 e^{-i \pmb{\mathcal{G}}_{2}( \pmb{\tau}_{\alpha} - \pmb{\tau}_{\beta})} 
 \braket{
 c^{(\nu)}_{\alpha,1;00}(\mathbf{k}_{a}) 
c^{\dag, (\nu)}_{\beta,2;00}(\mathbf{k}_{a} -\mathbf{q}) 
} 
\notag \\
& + J\left( \pmb{\mathcal{G}}_{2} -\pmb{\mathcal{G}}_{3} \right) 
e^{-i( \pmb{\mathcal{G}}_{2} - \pmb{\mathcal{G}}_{3}) \cdot ( \pmb{\tau}_{\alpha} - \pmb{\tau}_{\beta})} \braket{  
 c^{(\nu)}_{\alpha,1;00}(\mathbf{k}_{a}) 
c^{\dag, (\nu)}_{\beta,2;01}(\mathbf{k}_{a} -\mathbf{q})} 
\bigg\}
\end{align}

\enni \begin{align}
\tilde{t}^{ \alpha \beta)}_{I; n' m', n' m'+1} \left( \mathbf{k}, \mathbf{q} \right)
  = & 
\frac{4}{N} 
\sum_{\mathbf{k}_{a}} 
\bigg\{ 
  J\left(0 \right) 
 \braket{
 c^{(\nu)}_{\alpha,1;00}(\mathbf{k}_{a}) 
c^{\dag, (\nu)}_{\beta,2;01}(\mathbf{k}_{a} -\mathbf{q}) 
} 
+ J\left( \pmb{\mathcal{G}}_{3} \right)
 e^{-i \pmb{\mathcal{G}}_{3} \cdot( \pmb{\tau}_{\alpha} - \pmb{\tau}_{\beta})} 
 \braket{
 c^{(\nu)}_{\alpha,1;00}(\mathbf{k}_{a}) 
c^{\dag, (\nu)}_{\beta,2;00}(\mathbf{k}_{a} -\mathbf{q}) 
} 
\notag \\
& + J\left( -\pmb{\mathcal{G}}_{2} +\pmb{\mathcal{G}}_{3} \right) 
e^{i( \pmb{\mathcal{G}}_{2} - \pmb{\mathcal{G}}_{3}) \cdot ( \pmb{\tau}_{\alpha} - \pmb{\tau}_{\beta})} \braket{  
 c^{(\nu)}_{\alpha,1;00}(\mathbf{k}_{a}) 
c^{\dag, (\nu)}_{\beta,2;10}(\mathbf{k}_{a} -\mathbf{q})} 
\bigg\}
\end{align}

\enni \begin{align}
\tilde{t}^{(\alpha \beta)}_{II; n' m', n' m'} \left( \mathbf{k}, \mathbf{q} \right)
  = & 
\frac{4}{N} 
\sum_{\mathbf{k}_{a}} 
\bigg\{ 
  J\left(2 \mathbf{K}_{00} \right) 
 \braket{
 c^{\dag, (\nu)}_{\alpha,1;00}(\mathbf{k}_{a}) 
c^{(\nu)}_{\beta,2;00}(\mathbf{k}_{a} -\mathbf{q}) 
} 
\notag \\
+ & J\left(2 \mathbf{K}_{00} + \pmb{\mathcal{G}}_{2} \right)
 e^{-i \pmb{\mathcal{G}}_{2} \cdot( \pmb{\tau}_{\alpha} - \pmb{\tau}_{\beta})} 
 \braket{
 c^{ \dag, (\nu)}_{\alpha,1;00}(\mathbf{k}_{a}) 
c^{\dag, (\nu)}_{\beta,2;10}(\mathbf{k}_{a} -\mathbf{q}) 
} 
\notag \\
 + & J\left( 2\mathbf{K}_{00} + \pmb{\mathcal{G}}_{3}  \right) 
e^{-i \pmb{\mathcal{G}}_{3} \cdot ( \pmb{\tau}_{\alpha} - \pmb{\tau}_{\beta})} \braket{  
 c^{(\nu)}_{\alpha,1;00}(\mathbf{k}_{a}) 
c^{\dag, (\nu)}_{\beta,2;01}(\mathbf{k}_{a} -\mathbf{q})} 
\bigg\}
\end{align}

\enni \begin{align}
 \tilde{t}^{(\alpha \beta)}_{II; n' m', n'+1 m'} \left( \mathbf{k}, \mathbf{q} \right)
   =  &
\frac{4}{N} 
\sum_{\mathbf{k}_{a}} 
\bigg\{ 
  J\left(2 \mathbf{K}_{00} +2 \pmb{\mathcal{G}}_{2}\right) 
   e^{-2i \pmb{\mathcal{G}}_{2} \cdot( \pmb{\tau}_{\alpha} - \pmb{\tau}_{\beta})} 
 \braket{
 c^{\dag, (\nu)}_{\alpha,1;00}(\mathbf{k}_{a}) 
c^{(\nu)}_{\beta,2;10}(\mathbf{k}_{a} -\mathbf{q}) 
}
\notag \\ 
+ & J\left(2 \mathbf{K}_{00} + \pmb{\mathcal{G}}_{2} \right)
 e^{-i \pmb{\mathcal{G}}_{2} \cdot( \pmb{\tau}_{\alpha} - \pmb{\tau}_{\beta})} 
 \braket{
 c^{ \dag, (\nu)}_{\alpha,1;00}(\mathbf{k}_{a}) 
c^{\dag, (\nu)}_{\beta,2;00}(\mathbf{k}_{a} -\mathbf{q}) 
} 
\notag \\
 + & J\left( 2\mathbf{K}_{00} +\pmb{\mathcal{G}}_{2} + \pmb{\mathcal{G}}_{3}  \right) 
e^{-i ( \pmb{\mathcal{G}}_{2}+ \pmb{\mathcal{G}}_{3}) \cdot ( \pmb{\tau}_{\alpha} - \pmb{\tau}_{\beta})} \braket{  
 c^{(\nu)}_{\alpha,1;00}(\mathbf{k}_{a}) 
c^{\dag, (\nu)}_{\beta,2;01}(\mathbf{k}_{a} -\mathbf{q})} 
\bigg\}
\end{align}

\enni \begin{align} 
\tilde{t}^{(\alpha \beta)}_{II; n' m', n' m'+1} \left( \mathbf{k}, \mathbf{q} \right)
  = & 
\frac{4}{N} 
\sum_{\mathbf{k}_{a}} 
\bigg\{ 
  J\left(2 \mathbf{K}_{00} +2 \pmb{\mathcal{G}}_{3}\right) 
   e^{-2i \pmb{\mathcal{G}}_{3} \cdot( \pmb{\tau}_{\alpha} - \pmb{\tau}_{\beta})} 
 \braket{
 c^{\dag, (\nu)}_{\alpha,1;00}(\mathbf{k}_{a}) 
c^{(\nu)}_{\beta,2;01}(\mathbf{k}_{a} -\mathbf{q}) 
}
\notag \\ 
+ & J\left(2 \mathbf{K}_{00} + \pmb{\mathcal{G}}_{3} \right)
 e^{-i \pmb{\mathcal{G}}_{3} \cdot( \pmb{\tau}_{\alpha} - \pmb{\tau}_{\beta})} 
 \braket{
 c^{ \dag, (\nu)}_{\alpha,1;00}(\mathbf{k}_{a}) 
c^{\dag, (\nu)}_{\beta,2;00}(\mathbf{k}_{a} -\mathbf{q}) 
} 
\notag \\
 + & J\left( 2\mathbf{K}_{00} +\pmb{\mathcal{G}}_{2} + \pmb{\mathcal{G}}_{3}  \right) 
e^{-i ( \pmb{\mathcal{G}}_{2}+ \pmb{\mathcal{G}}_{3}) \cdot ( \pmb{\tau}_{\alpha} - \pmb{\tau}_{\beta})} \braket{  
 c^{(\nu)}_{\alpha,1;00}(\mathbf{k}_{a}) 
c^{\dag, (\nu)}_{\beta,2;10}(\mathbf{k}_{a} -\mathbf{q})} 
\bigg\}
\end{align}

\enni Finally, these expressions simplify considerably once we ignore the relative variation of the different $J$'s, and we recover the form discussed in the main text. 

\section{Mean-field procedure}

The MF parameters $\braket{ c^{\dag, (\mu)}_{1, \alpha; nm}(\mathbf{k}) 
c^{(\mu)}_{2, \beta;nm}(\mathbf{k} -\mathbf{q})}$
 are defined for $\mathbf{q}=0$ for finite $\mathbf{q}$ in Eqs. 7 and 8 of the main text. Here, 1 and 2 are the layer indices, $\alpha, \beta$ are the sublattice indices, $\mu$ denote the three Majorana flavors, and $n,m$ label the Dirac points in the extended moire Brillouin zone. These are determined for $(\alpha, \beta) \in \{\text{(A,A), (A, B), (B, A), (B, B)}\}$ with $(n,m) \in \{(0,0),(1,0),(0,1) \}$ for NN Dirac points. Our solutions are chosen to the preserve the SO(3) symmetry of the model and are thus independent of $\mu$. There are therefore 12 MF parameters, each of which is determined without imposing any additional conditions. The calculations were performed in an extended moire Brillouin zone covering 100 unit cells. 

\section{Topological ground-state degeneracy in the mean-field approximation}

In this section, we demonstrate the topological degeneracy of the GS manifold, as determined from the Hartree approximation. In the following, we assume an even number of unit cells along both directions of the Bravais lattice. 

In Eq. 4 of the main text, we mapped the Yao-Lee bilayer with AA stacking onto a Hubbard model with three flavors of complex fermions. The mapping assumed uniform bonds for both layers with $u^{(\alpha)}_{1, ij} = u^{(\alpha)}_{2, ij}$ for all $\alpha, i$, and $j$. Here, we generalize this procedure, by choosing identical bond variables for both layers 

\enni \begin{align}
u^{(\alpha)}_{1, ij} = u^{(\alpha)}_{2, ij} = 
u^{(\alpha)}_{ij},
\end{align}

\enni while still allowing $u^{(\alpha)}_{ij}= \pm 1$. 
This allows us to consider arbitrary, fixed $u^{(\alpha)}_{ij}$ corresponding to topologically distinct sectors. The Yao-Lee bilayer is mapped onto a single-layer Hubbard model with three flavors of complex fermions coupled to a $\mathbb{Z}_{2}$ gauge field: 

\begin{eqnarray}
     \mathcal{H}_{c} =2K \sum_{\langle ij \rangle, \alpha}(i u^{(\alpha)}_{ij} f_{\rm{A},i}^{\alpha \dagger}f_{\rm{B},j}^{\alpha} +{\rm H.c.}) -2J\sum_{i}\left(n_i-\frac{3}{2}\right)^2.
\end{eqnarray}

\enni $\mathcal{H}_{c}$ is invariant under simultaneous gauge transformations on both layers

\enni \begin{align}
D'_{i} = & D_{1, i} D_{2, i},
\end{align}

\enni where the $D_{1/2, i}$ operators were defined in the main text. $D'_{i}$ maps $f^{(\alpha)}_{i} \rightarrow - f^{(\alpha)}_{i}$ and $u^{(\alpha)}_{\braket{ij}} \rightarrow - u^{(\alpha)}_{\braket{ij}}$. Note that for a given, non-trivial gauge choice, the physical GS is still obtained via application of the projection operator $P$.

We recall that 

\enni \begin{align}
\braket{\chi^{(\alpha)}_{i}} 
=  \braket{2 n^{(\alpha)}_{i} - 1},
\end{align}

\enni was introduced in the main text for the gauge with all $u^{(\alpha)}_{ij}=1$. This MF parameter is invariant under the gauge transformations implemented by $D'_{i}$. For the more general cases considered here, we introduce similar parameters

\begin{align}
\mathcal{H}'_{c} =2K \sum_{\langle ij \rangle, \alpha}(i u^{(\alpha)}_{ij} f_{\rm{A},i}^{\alpha \dagger}f_{\rm{A},j}^{\alpha} +{\rm H.c.}) 
+ m_{\rm{A}} \sum_{i \in \rm{A}} n_{\rm{A},i}
+ m_{\rm{B}}  \sum_{i \in \rm{B}} n_{\rm{B},i},
\label{Eq:Hcmf}     
\end{align}

\enni where we neglected a trivial shift in energy. The MF parameters for gauge choice $\{ u \}$ are 

\enni \begin{align}
m_{\rm{A/B}} = -8J \sum_{\alpha} \braket{\chi^{(\alpha)}_{\rm{A/B}}}_{\{u \} }.
\label{Eq:morp}
\end{align}

In order to classify the topological degeneracy of the GS manifold of $\mathcal{H}'_{c}$, we consider two Wilson loop operators~\cite{S_Chulliparambil_PRB2020}

\enni \begin{align}
W_{1/2} = \Pi_{\alpha -\text{links}, \braket{ij}~\in C_{1/2}}~~u^{(\alpha)}_{ij}
\end{align}

\begin{figure}[h!]
\includegraphics[width=0.5\columnwidth]{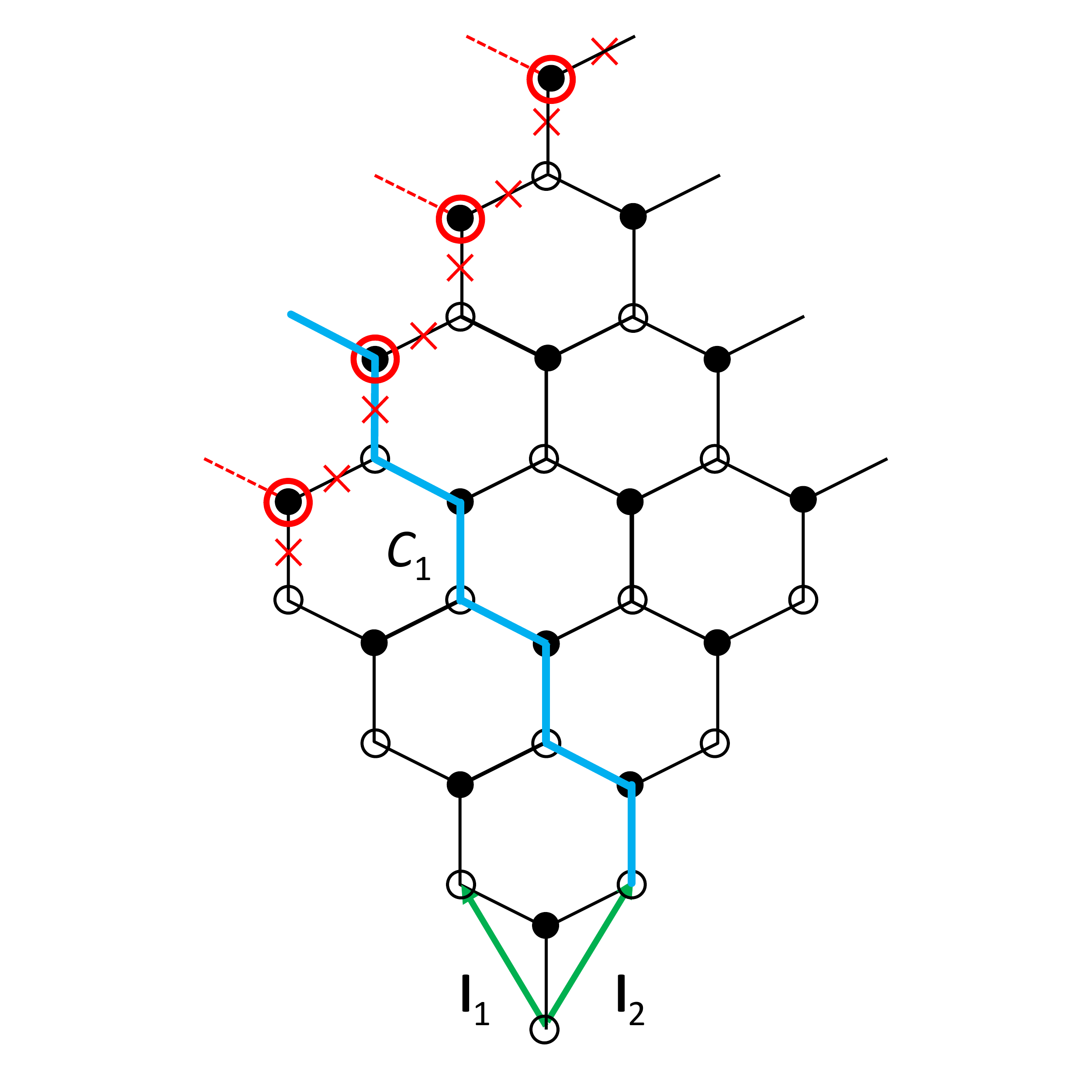}
\caption{Illustration of the loop $C_{1}$, marked in blue, corresponding to the Wilson operator $W_{1}$. The green arrows indicate the two Bravais lattice vectors $\mathbf{l}_{1/2}$. All unmarked bonds are assumed to take values $u^{(\alpha)}_{ij}=1$. The bonds marked with a red cross, which stand for  $u^{(\alpha)}_{ij}=-1$, form a non-contractible loop. It can be made equivalent to anti-periodic boundary conditions (APBC) along $\mathbf{l}_{2}$ by applying the $D'$ operators on sites marked with red circles. These bond configurations correspond to eigenstates of $W_{1}$ with eigenvalue $\lambda_{1}=-1$. 
Similar arguments hold for $W_{2}$.}
\label{Fig:Wllp}
\end{figure}

\enni where $C_{1/2}$ are non-contractible loops along the two Bravais lattice vectors $\mathbf{l}_{1/2}$, as illustrated in Fig.~\ref{Fig:Wllp}. Since $W_{1/2}$ commute with $\mathcal{H}'_{c}$ and $W^{2}_{1/2}=1$, we can label the eigenstates of the Hamiltonian via the $\lambda_{1/2} = \pm 1$ eigenvalues of the two loop operators. 

Configurations of the bond variables with contractible loops are gauge-equivalent to the uniform bond configuration where $u^{(\alpha)}_{ij}=1$ for any $i,j$ NN pairs. With this choice of gauge, $\mathcal{H}'_{c}$ obeys periodic boundary conditions (PBC) along both directions of the Bravais lattice. As already discussed in the main text, here $m_{\rm{A}} = - m_{\rm{B}} = m$, and the GS energy is 

\enni \begin{align}
E_{\rm{GS}} = & - 3 \sum_{\mathbf{k}} \sqrt{ \left| 2K f(\mathbf{k}) \right|^{2} + m^{2}},
\end{align}

\enni where 

\enni \begin{align}
f(\mathbf{k}) = &  e^{i\mathbf{k} \cdot(\pmb{\tau}_{A} - \pmb{\tau}_{B})}
\left(1+
e^{-i\mathbf{k} \cdot \mathbf{l}_{1}}
+ 
e^{-i\mathbf{k} \cdot \mathbf{l}_{2}}
\right),
\end{align}

\enni $\pmb{\tau}_{A/B}$ are the positions of the two sublattice sites in the unit cell, and $\mathbf{l}_{1/2}$ are the two primitive Bravais lattice vectors, as illustrated in Fig.~\ref{Fig:Wllp}. Note that $ \left| f(\mathbf{k}) \right|^{2}$ is invariant under translation by the reciprocal unit vectors $\pmb{\mathcal{G}}_{2/3}$ defined previously. 

Next, we consider bond configurations with non-contractible loops. In Fig.~\ref{Fig:Wllp} we illustrate one such loop along the $\mathbf{l}_{2}$ direction, where the bonds with $u^{(\alpha)}_{ij}=-1$ are marked by red crosses. This configuration can be labeled by $\lambda_{1}=-1, \lambda_{2}=1$. It is also equivalent to another configuration obtained by flipping the bonds marked with red, dashed lines to negative values, while  setting all other bonds to be positive. The two configurations are transformed into each other  by applying gauge transformations $D'_{i}$ at every site marked by a red circle in Fig.~\ref{Fig:Wllp}. Consequently, the presence of the non-contractible loop along $\mathbf{l}_{2}$ is equivalent to adopting anti-periodic boundary conditions (APBC) along $\mathbf{l}_{1}$. It follows that configurations corresponding to $\lambda_{1}=1, \lambda_{2}= -1$ and $\lambda_{1}=-1, \lambda_{2}= -1$ can be similarly constructed by adopting PBC/APBC and APBC/APBC along $\mathbf{l}_{1/2}$, respectively.

We implement APBC along $\mathbf{l}_{1/2}$ by shifting  the primitive reciprocal unit cell for PBC by  $\pmb{\mathcal{G}}_{2/3}/ 2N_{1/2}$, where $N_{1/2}$ are the numbers of unit cells along either direction. The on-site parameters are invariant under these shifts, ensuring that we recover the results for PBC with $m_{\rm{A}} =- m_{\rm{B}} = m $. Consequently, all four distinct topological GS sectors with $\lambda_{1/2} = \pm 1$ are degenerate, in agreement with our conclusions based on the perturbative analysis in the large-$J$ limit. Finally, we note that all four GS topological sectors survive projection. As shown in Eqs.~\ref{Eq:Prjc} and~\ref{Eq:Prty}, the projection operator depends on the products of bonds in both layers, and on the total, complex-fermion parity. Our conclusion is due to the choice of identical bonds in both layers $u^{(\alpha)}_{1,ij}= u^{(\alpha)}_{2,ij}$, and of the half-filling of the complex fermions, as previously discussed in Sec.~\ref{Sec:Prjc}.

\section{Effect of spin operators in Kitaev and Yao-Lee bilayers}

In this section, we contrast the effects of the spin operators in Kitaev and Yao-Lee models. We first consider the Kitaev model and adopt the Majorana representation of Ref.~\onlinecite{S_Kitaev_AnnPhys2006} for the spin operators

\enni \begin{align}
\sigma^{(\alpha)}_{i} = & i b^{(\alpha)}_{i} c_{i},
\end{align}

\enni together with the constraint

\enni \begin{align}
D_{i} = &  b^{(x)}_{i} b^{(y)}_{i} b^{(z)}_{i} c_{i} 
\notag \\
= & 1
\end{align}

\enni one every site $i$. We next consider the bond operators 

\enni \begin{align}
\hat{u}^{(\alpha)}_{ij} = i b^{(\alpha)}_{i} b^{(\alpha)}_{j}.
\end{align}

\enni The spin operators anti-commute with the bond operators

\enni \begin{align}
\{\sigma^{(\beta)}_{k}, \hat{u}^{(\alpha)}_{ij} \} = & (\delta_{ki}+ \delta_{kj}) \delta_{\alpha \beta}.
\end{align}

\enni Consequently, we can write the following

\enni \begin{align}
\braket{\hat{u}^{(\alpha)}_{ij}} 
= - \braket{ \sigma^{(\alpha)}_{i} \hat{u}^{(\alpha)}_{ij} \sigma^{(\alpha)}_{i}}, 
\end{align}

\enni which indicates that $\sigma^{(\alpha)}_{i}$ flips the bond $\alpha$ at vertex $i$, and therefore creates two visons. 

By contrast, $\sigma^{(\alpha)}_{i}$ commutes with $\hat{u}^{(\alpha)}_{ij}$ in the Yao-Lee model, since these operators are expressed in terms of $c$ and $b$ Majorana fermions, respectively, as shown in the main text. Consequently, the spin operators here preserve the flux.


\end{document}